\documentclass[nonatbib,noprefixes]{cc}
\newcommand{\PP}{\mathbb{P}}
\newcommand{\R}{\mathbb{R}}
\newcommand{\Q}{\mathbb{Q}}

\newcommand{\K}{\mathbb{K}}
\newcommand{\Z}{\mathbb{Z}}
\newcommand{\LL}{\mathbb{L}}
\newcommand{\F}{\mathbb{F}}
\newcommand{\CF}{\overline{\F}}
\newcommand{\CK}{\overline{\K}}
\newcommand{\CZ}{\overline{Z}}

\newcommand{\eps}{\varepsilon}
\newcommand{\Le}{\LL\langle\eps\rangle}
\newcommand{\ea}{\langle\eps\rangle}
\newcommand{\Ke}{\K\ea}
\newcommand{\Fe}{\F\ea}
\newcommand{\CFe}{\CF\ea}
\newcommand{\CFz}{\CF\langle\zeta\rangle}

\newcommand{\Fz}{\F\langle\zeta\rangle}

\newcommand{\qq}{q}

\newcommand{\tp}{{\tilde{p}}}
\newcommand{\tQ}{{\tilde{Q}}}
\newcommand{\tk}{{\tilde{k}}}
\newcommand{\tn}{{\tilde{n}}}

\newcommand{\beq}[1]{\begin{equation} \label{eq:#1} }
\newcommand{\refeq}[1]{(\ref{eq:#1})}
\newcommand{\eeq}{\end{equation}}

\newcommand{\sign}{\mathrm{sign\, }}
\newcommand{\diag}[1]{\mathrm{diag}({#1})}
\newcommand{\rank}[1]{\mathrm{rk}({#1})}
\newcommand{\Tr}[1]{\mathrm{Tr}({#1})}

\contact{d.v.pasechnik@uvt.nl}
\received{March 2, 2004}

\title{Polynomial-time computing over quadratic maps I: \\
sampling in real algebraic sets.}

\titlehead{Computing over quadratic maps I} 

\author{
Dima Grigoriev\\
IRMAR\\
Universit\'e de Rennes I\\ 
Campus de Beaulieu\\
35042 Rennes cedex\\ 
France\\
\email{dima@math.univ-rennes1.fr}\\
\homepage {http://name.math.univ-rennes1.fr/dimitri.grigoriev/}
\and
Dmitrii V. Pasechnik\\
Dept. E \& OR and CentER\\
Tilburg University\\
P.O. Box 90153\\
5000 LE Tilburg\\
The Netherlands\\
\email{d.v.pasechnik@uvt.nl}\\ 
\homepage{http://center.uvt.nl/staff/pasechnik/}
}

\begin{abstract}
Given a quadratic map $Q : \K^{n}\rightarrow \K^k$ defined over a
computable subring $D$ of a real closed
field $\K,$ and $p\in D[Y_1,\dots,Y_k]$ of degree $d$ 
we consider the zero set
$Z=Z(p(Q(X)),\K^n)\subseteq\K^n$
of $p(Q(X_1,\dots,X_n))\in D[X_1,\dots,X_n]$. 
We present a procedure that computes, in $(dn)^{O(k)}$ arithmetic
operations in $D,$ a set $\mathcal{S}$
of (real univariate representations of) sampling
points in $\K^n$ that intersects nontrivially each connected component
of $Z.$ As soon as $k=o(n),$
this is faster than the standard methods %(see e.g. \cite{GV88,BPR03}) 
that all have exponential dependence on $n$ in the complexity. 
In particular, our procedure is polynomial-time for constant $k$.
In contrast, the best previously known procedure %\cite{Bar93}
is only capable of deciding in $n^{O(k^2)}$ operations the
nonemptiness (rather than constructing sampling points)
of the set $Z$ in the case of $p(Y)=\sum_i Y_i^2$
and homogeneous $Q$.

A by-product of our procedure is a bound $(dn)^{O(k)}$ on the 
number of connected components of $Z$.

The procedure consists of exact symbolic computations in $D$
and outputs vectors of algebraic numbers.
It involves extending $\K$ by infinitesimals and subsequent
limit computation by a novel procedure
that utilizes knowledge of an explicit isomorphism
between real algebraic sets.
\end{abstract}


\begin{subject}
68W30, 13P10, 14Q20, 14Pxx
\end{subject}

\begin{document}
%\maketitle
\section{Introduction and the results} 
The algorithmic problem of finding points in real algebraic 
sets has received considerable 
attention, in particular as 
it forms a building block for a lot of procedures in real 
algebraic geometry \cite{BPR03}. Even if the algebraic sets one is 
interested in are subsets of $\R^n$,
the algorithms with the best known complexity bounds 
use transcendental infinitesimals extending $\R$ to perform 
necessary geometric deformations of the sets. Hence it is 
natural to describe the procedures as operating over an arbitrary 
real closed field $\K$, with the input data, i.e. the polynomials, 
lying in $D[X_1,\dots,X_n]=D[X]$, with $D\subset\K$ a computable 
(in sense discussed e.g. in \cite[Sect.~8.1]{BPR03}) subring 
of $\K.$ In the case $\K=\R$ one usually assumes $D=\Z.$ 

Let $\mathcal{S}_i$'s be the  connected components 
of the real algebraic set $\mathcal{S}=Z(f,\K^n),$ $f\in D[X]$.
%(i.e. the set of roots of $f$ in $\K^n$).
In general the number of the $\mathcal{S}_i$'s is bounded by $d^{O(n)}$, where 
$d=\deg f$, and this bound is sharp, see 
\cite[Theorem~7.23, Remark~7.22]{BPR03} and Remark~\ref{rem:hypercube}
below. 
We are interested in point-finding (also called {\em sampling}) 
algorithms that  
produce a finite set of points that intersects each $\mathcal{S}_i$. 
Such algorithms with the best known complexity bounds 
need at most $d^{O(n)}$ arithmetic operations in $D$.
Often such sets are exactly what is needed in 
applications. We describe  here a procedure 
that finds a point in each connected component of our class 
of algebraic sets, namely the sets of the form $Z(p(Q(X)),\K^n)$,  
for $p\in D[Y_1,\dots,Y_k]$ of degree $\deg p=d$ and 
$Q=(Q_1(X),\dots,Q_k(X))\in D[X]^k$ a quadratic map, i.e. 
$\deg Q_j\leq 2$ for $1\leq j\leq k$ with the complexity
$(dn)^{O(k)}$.

The result of \cite{Bar97}, 
that bounded, in particular, 
the sum of the Betti numbers of the  set of real solutions of 
a system of quadratic equations $Q_1(X)=\dots =Q_k(X)=0$ 
(that can obviously be written as 
$Z(p(Q(X)),\R^n)$ with $p(Y)=\sum_j Y_j^2$) by a 
polynomial in $k$ and $n$ of degree $O(k)$ was perhaps the 
earliest indication that in the case $Z(p(Q(X)),\K^n)$ the 
number of connected components has only polynomial dependence 
on $n.$ However, until the present work, an algorithmic procedure with 
the similar complexity bound $(dn)^{O(k)}$ for 
finding points in $Z(p(Q(X)),\K^n)$ was unknown. 
Even procedures that decide non-emptiness of 
$Z(p(Q(X)),\K^n)$ in time polynomial in $n$ and $d$ for fixed $k$ 
were, for general $p,$ unknown; in \cite{Bar93} such a procedure 
was described for $p(Y)=\sum_j Y_j^2$, homogeneous $Q_i$'s 
and $\K=\R$. 

The technique we use is that of symbolic computation. All the 
data is represented exactly, as (real) algebraic numbers, if necessary. 
More precisely, elements of $\K^n$ that we compute with are 
given by real univariate representations. The latter are defined as follows. 
A {\em sign condition} 
for a set of polynomials $\mathcal{P}=\{P_1,\dots,P_s\}\subset \K[Y]$ 
is specified by $\sigma\in \{-1,0,1\}^s$ 
so that $\sigma=(\sign{P_1(Y)},\dots,\sign{P_s(Y)})$. 
\emph{Thom encoding} \cite[Lemma~2.38,~Sect.~10.4]{BPR03} 
of a root $\alpha\in\K$ of $f\in\K[T]$ is a sign condition $\sigma_\alpha$ 
on the derivatives of $f$, that is 
$\sigma_\alpha=(\sign{f'(\alpha)},\dots,\sign{f^{(\deg{f}-1)}(\alpha)})$. 
Note that $\sigma_\alpha$ and $f$ determine $\alpha\in\K.$ 
Let $\CK$ denote the algebraic 
closure of $\K.$ 
A {\em univariate representation} of 
$u\in\CK^m$ is an $(m+2)$-tuple 
\beq{def:univar}
u(T)=(f,g_0,g_1,\dots,g_m)
\eeq
of univariate polynomials in $D[T]$ satisfying 
$u=\frac{1}{g_0(\alpha)}(g_1(\alpha),\dots,g_m(\alpha))$ for a root 
$\alpha\in\CK$ 
of $f$, and such that $f$ and $g_0$ are coprime. 
Obviously, each $u(T)$ can represent 
as many as $\deg(f)$ distinct elements of $\CK^m.$ 
A {\em real} univariate representation of $u\in\K^m$ is a 
pair $u(T), \sigma_\alpha$, where $\sigma_\alpha$ is the Thom encoding 
of a root $\alpha\in\K$ of $f.$ 

The main result of the paper is as follows. 
\begin{theorem}\label{thm:mainthm_pQ} 
Let $Q=(Q_1,\dots,Q_k)\in D[X_1,\dots,X_n]^k$ be a quadratic map 
$X\mapsto Q(X),$ and
let $p\in D[Y_1,\dots,Y_k]$ satisfy $\deg p\leq d$.
A set of real univariate representations $u(T),\sigma_\alpha$ 
of a set of points in $Z=Z(p(Q(X)),\K^n)$ meeting each connected component of 
$Z$ can be computed in $(dn)^{O(k)}$ arithmetic 
operations in $D$. 
The degrees of polynomials in $u(T)$ are bounded by $(dn)^{O(k)}$.
When $D=\Z$, the coefficients of $u(T)$ and the intermediate 
polynomial data will be bounded by  $(dn)^{O(k)}$ 
times the bitsize of the input data $p$, $Q$. 
\end{theorem} 

From now on whenever we talk about finding points in $\K^n$, they 
are meant to be given as real univariate representations. 

\begin{remark}
When $\K$ is archimedean, e.g. $\K=\R$, 
the approximations, in the ring of fractions of $D$, of the point in $\K^n$ 
given by a real univariate representation 
can be found efficiently as long as approximations
$\tilde{\alpha}$ of  $\alpha$ can be computed 
efficiently (indeed, then one can just compute $u(\tilde{\alpha})$).  
For instance when $D=\Z$ one can find an interval 
$\mathcal{I}=[\alpha_-,\alpha_+]\ni\alpha$ with $\alpha_\pm\in\Q$ so that
$\alpha$ is the only root of $f$ in $\mathcal{I}$, see
e.g. \cite[Sect.~10.2]{BPR03}. Once $\mathcal{I}$ is known,
one can compute its repeated (rational) 
bisections to obtain approximations of 
$\alpha$ of needed precision; the complexity of the latter is
analyzed e.g. in \cite{MR45:9480} (see also \cite{GaGe99}).
\end{remark}

Note that by {\em connected} (component of) semialgebraic set, 
we mean {\em semialgebraically connected}, that is, 
connected in the semialgebraic topology, (component of) semialgebraic set, 
see e.g. \cite{Gri88}. It is well-known that for the semialgebraic 
sets over $\R$  semialgebraic connectedness implies connectedness 
(in the usual Euclidean topology), see e.g. \cite[Thm.~5.21]{BPR03}. 

Theorem~\ref{thm:mainthm_pQ} is proved in Section~\ref{sect:mainproof} by 
exhibiting a procedure that does the claimed task.
It immediately implies the following.
\begin{corollary} \label{cor:numconcom}
The number of connected components of the set 
$Z$ %=Z(p(Q(X)),\K^n)$ 
is at most $(dn)^{O(k)}$. \qed 
\end{corollary} 
An extra argument, to be published elsewhere, allowed us to show that
the latter bound holds for the sum of Betti numbers of $Z$, and
not only for the $0$-th one, i.e. the number of components. As well,
one can modify the procedure of Theorem~\ref{thm:mainthm_pQ} to prove 
\begin{theorem}\label{thm:optthm_pQ} 
Under the assumptions of Theorem~\ref{thm:mainthm_pQ}, 
computing the exact 
minimum and a minimizer (i.e. a point where the minimum is attained)
of $r(Q(X)),$ for $r\in D[Y],$ $\deg r\leq d,$ 
on $Z(p(Q(X)),\K^n),$ or checking that the minimum is not 
attained and computing the infimum, 
can be done within the same number of operations, 
and for $D=\Z$ within the same bitwise complexity, as the 
computation of Theorem~\ref{thm:mainthm_pQ}. 
\end{theorem} 
A proof of the latter, and a number of applications in mathematical 
programming, will appear in the continuation of the 
present paper.

An easier than optimization problem is the problem of checking 
whether the set $Z(p(Q(X)),\K^n)$ is empty, i.e. the feasibility 
problem. Our immediate predecessor here is \cite{Bar93}, where 
it was shown that for homogeneous $Q$ the emptiness 
of $Z(Q_1(X)^2+\dots+Q_k(X)^2,\R^n-\{0\})$ can be checked in $n^{O(k^2)}$ 
operations in $D.$ 

For the sake of completeness, we state the following 
straightforward implication of Theorem~\ref{thm:mainthm_pQ}. 
\begin{corollary} 
The emptiness of $Z(p(Q(X)),\K^n)$ can be checked 
within the same complexity bound as in  
Theorem~\ref{thm:mainthm_pQ}. \qed 
\end{corollary} 

\begin{remark}\label{rem:hypercube}
It is easy to see that the bounds of Theorem~\ref{thm:mainthm_pQ} 
are close to best possible. 
Indeed, any real solution of degree 4 equation
$$(X_1^2-1)^2+\dots+(X_n^2-1)^2=0,$$
or the system of $n$ quadratic equations
$$X_1^2=X_2^2=\dots=X_n^2=1$$
has coordinates $1$ or $-1$, and there are in total $2^n$ of them.
In the continuation of the present paper 
we will further sharpen this by showing
a similar result for a system of one cubic and two quadratic equations.
\end{remark}

In a nutshell, the procedure 
at the core of Theorem~\ref{thm:mainthm_pQ} that 
we are going to describe works as follows.
First, we write down the equations for the critical points of the
projection map $X\mapsto X_1$ on  $Z=Z(p(Q(X)),\K^n)$ by equating
the gradient of $p(Q(X))$ with the vector proportional to
the gradient of $X\mapsto X_1,$ that is with a vector of
the form $(\lambda,0,\dots,0).$ 
These equations have a rather special structure: the variables $X$
occur either within $Q(X)$, or linearly.
By introducing $k$ new variables $(Y_1,\dots,Y_k)=Y=Q(X)$, 
we thus obtain a system of
{\em linear} equations $A(Y)X=b(Y)$ in $X.$
The next step is to solve this system; we simply loop through all the
maximal (by inclusion) candidates for invertible submatrices $A_{UW}$ 
of $A(Y)$ and the corresponding partition 
$X_{W}\cup X_{\overline{W}}$ of $X$ into $X_W$ and 
the remaining variables $X_{\overline{W}}.$ For each of them we
rewrite the system to express $X_W$ as 
rational functions $X_W=A'(Y,X_{\overline{W}})$
of $Y$ and $X_{\overline{W}}.$
This certainly only makes sense, from the complexity point of view, 
when $\rank{A(Y)}$ never 
drops below certain threshold. We  make sure by
means of an infinitesimal deformation that
$\rank{A(Y)}\geq n-k$. Then $|\overline{W}|\leq k$ and the coordinates
of $X$ are expressed as rational functions in at most $2k$ variables
$Y$ and $X_{\overline{W}}.$ 
We are able to describe
an isomorphism of a semialgebraic subset, that we call,
following \cite{Bar93},  \emph{piece}, of the critical points 
of $X\mapsto X_1$ on $Z,$ that corresponds
to a particular $A_{UW}$ being maximal and invertible, to 
a semialgebraic subset of $\F^{k+|\overline{W}|}$, for $\F$ 
being a real closed extension of $\K,$ that is
defined by polynomials of degree $O(nd).$ These pieces cover the
whole set of the critical points just mentioned.

Finally, we find representatives of connected components of
the pieces over $\F$,
obtaining $Y$ and $X_{\overline{W}}$ with values in $\F$, 
and recover $X_W$ and 
$X_{\overline{W}}$ in the original field $\K$ by 
computing the limit.

The actual implementation of this procedure is more involved. 
Section~\ref{sect:pieces} describes in detail the candidates for 
invertible submatrices $A_{UW}$ of $A$ and presents  
the explicit isomorphisms of pieces to semialgebraic
subsets in $\F^{k+|\overline{W}|}$ mentioned above. 
In order to apply the result of Section~\ref{sect:pieces} to $Z$,
%=Z(p(Q(X)),\K^n),$ 
one needs to deform $p$ in such a way
that $0$ becomes a regular value of $p(Q(X))$ and of $p(Y)$. Further,
one needs to deform $Q$ so that the number of pieces of the set of 
critical points of $X\mapsto X_1$ on $Z$ does not exceed $(dn)^{O(k)}.$
In fact, our deformation will give us a better bound, $n^{O(k)},$ 
on the latter. 
Lastly, one has to ensure (again, using a deformation)
that $Z$ is bounded, otherwise we 
miss connected components of $Z$ whose projection on $X_1$ is 
open.

Our deformations are done by extending $\K$ with a number of infinitesimals.
Subsequent limit computations are needed to recover elements in the original
set $Z$ by using the following Theorem~\ref{thm:imlim}.
To state it, let us recall some notation.
For a field $\F$ and a transcendental $\zeta$, we denote
by $\Fz\subset\F((\zeta^\frac{1}{\infty}))$ the subfield of 
Puiseux series algebraic over $\F(\zeta)$.
For 
\beq{puis}
a=\sum_{i\geq\nu} a_i\zeta^{i/q}\in\F((\zeta^\frac{1}{\infty})),\qquad 
0<q\in\Z
\eeq
with the order $\nu/q\geq 0$, $a_\nu\neq 0$, 
define the {\em standard part} (cf. e.g. \cite{GV88}) 
of $a$ to be $a_0$; in \cite{BPR03} it is called the {\em limit}
$a_0=\lim_{\zeta}a$. 
Note that if $\nu<0$ then $\lim_\zeta a$ is not defined. 
When $\zeta$ is a vector of infinitesimals 
$\zeta_1\gg\zeta_2\gg\dots\gg\zeta_\ell$, 
the notation $\lim_\zeta a$ is a shorthand for 
$\lim_{\zeta_1} (\lim_{\zeta_2} (\dots (\lim_{\zeta_\ell} a)\dots ))$.
It is often helpful to view $\zeta$ as a {\em parameter} and computing
$\lim_{\zeta}a$ as computing $\lim_{\zeta\rightarrow 0}a$, where $\lim$ is
understood in the usual sense. Note that $\lim_\zeta$ is 
a ring homomorphism of the ring $\Fz_b=\{a\in\Fz\mid \nu(a)\geq 0\},$
of all the elements of $\Fz$ \emph{bounded} over $\F$, to $\F$.

Let $\Fe$ be a real closed extension of a real closed 
$\F$ with infinitesimals $\eps=(\eps_1,\dots,\eps_\ell)$ such that
$\eps_1\gg\eps_2\gg\dots\gg\eps_\ell$, and let $D\subset\F$ be a computable
subring of $\F$.
For $F\in D[\eps][Y_1,\dots,Y_{\qq-1}]$, 
let  $Z_F=Z(F(Y),\Fe^{\qq-1})$, and let 
\begin{equation}\label{def:Psi}
\begin{split}
\Psi:\Fe^{\qq-1}&\rightarrow  \Fe^m\qquad\text{be a rational mapping}\\ 
Y&\mapsto  \left(\frac{\Omega_1(Y)}{\Lambda(Y)},\dots, 
  \frac{\Omega_m(Y)}{\Lambda(Y)}\right),\quad 
  \Omega_i,\Lambda\in D[\eps][Y],\quad 1\leq i\leq m.
  \end{split}
\end{equation}

\begin{theorem}\label{thm:imlim}
Let $F$ and $\Psi$ be as above, with the $Y$-degree of $F$ at most $d$
and the $Y$-degrees of $\Omega_i$ and $\Lambda$ less than $d-1$, and
their $\eps$-degrees at most $d$.
A set of univariate representations $u(T)$ 
of a set of points meeting each connected component of 
$\lim_\eps\Psi(Z_F)\subseteq \F^m$ 
can be computed in 
$(m+d)^{O(\qq\ell)}$ arithmetic operations in $D$.\\
The degrees of the polynomials in $u(T)$ are at  
most $d^{O(\qq)}$. 
When $D=\Z$, the bitsizes of the coefficients of 
$u(T)$ and of the intermediate data
are bounded by a polynomial in $d$, $m$ and $d^{O(\qq\ell)}$ 
times the bitsize of the input data.
\end{theorem}
Theorem~\ref{thm:imlim} generalizes \cite[Alg.~11.61]{BPR03}
to non-identity mappings $\Psi$.

The remainder of the paper begins 
with presenting the procedures behind Theorem~\ref{thm:imlim},
along with its proof, in Sections~\ref{sect:limimagedim0}
and \ref{sect:limimage}. 
Then Section~\ref{sect:pieces} presents the aforementioned decomposition
of the zero set of $p(Q(X))$ under the regularity conditions.
Finally, Section~\ref{sect:mainproof} describes the deformations
of $p(Q(X))$ that are needed and completes the proof of the main 
Theorem~\ref{thm:mainthm_pQ}. 

\section{Limits of solution images: dimension 0}\label{sect:limimagedim0}
\newcommand{\mcB}{\mathcal{B}}
\renewcommand{\PP}{P}
\newcommand{\gpolhat}[3]{\hat{g}(#1,#2,#3)}
\newcommand{\gpolhatd}[4]{\hat{g}^{({#1})}(#2,#3,#4)}
\newcommand{\charpol}[2]{\chi({#1},{#2})}
\newcommand{\gpol}[3]{g({#1},{#2},{#3})}
\newcommand{\gpold}[4]{g^{({#1})}({#2},{#3},{#4})}
\newcommand{\mul}[1]{{\mu_{[{#1}]}}}
\newcommand{\mux}{{\mul{x}}}
\newcommand{\mus}{{\mul{s}}}
\newcommand{\muy}{{\mul{y}}}
\newcommand{\In}[1]{\mathrm{in}({#1})}
\newcommand{\Gf}{\mathfrak{G}}
As the first part of the proof of Theorem~\ref{thm:imlim}, 
in this section 
we address the problem of  finding limits of the images $\PP(x)$ of
the real roots $x\in Z(\mcB,\Fe^\qq)$ of a 0-dimen\-sional polynomial system 
$\mathcal{B}\subset\Fe^\qq$ 
with respect to $\eps=(\eps_1,\dots,\eps_\ell)\rightarrow 0$ 
under a polynomial mapping
$\PP:\Fe^\qq\rightarrow\Fe^m$.

The separating element based methods for finding limits of 
the real roots of  $\mathcal{B}$, such as
\cite{RRS00}, \cite[Chapter~11]{BPR03}, 
cannot deal directly with this
situation, %.  Indeed, the roots themselves need not be bounded, 
%and may approach infinity when an 
%infinitesimal approaches 0. That is, 
even in the simplest case of 
$\PP$ being the orthogonal  projection onto a subset of coordinates.
%one needs a generalization of these methods
%to deal with the problem. 

In this section we generalize these methods to accommodate our needs.
%Let $A=\CFe[X]/(\mcB)$ be
%the quotient algebra of $\CFe[X]$ over the ideal 
%$(\mcB)\subseteq\CFe[X]$.
We introduce $\PP$-\emph{separating elements} 
$a\in\CFe[S_1,\dots,S_m]$
such that the map $\PP(x)\mapsto a(\PP(x))$ is injective on $\PP(\CZ),$ 
where $\CZ=Z(\mcB,\CFe^\qq),$ 
that is to say that $a(\PP(y))\neq a(\PP(x))$ whenever $\PP(y)\neq\PP(x)$ for
$x,y\in\CZ$.
We introduce below, in Section~\ref{subs:comli},  
the corresponding notion for the limit
setting, \emph{well-$\PP$-separating elements}.

\newcommand{\maW}{\mathcal{W}}
\newcommand{\oU}{\overline{\mathrm{U}}}

It turns out that the machinery of \cite{RRS00}, see also \cite{ABRW96}, 
generalizes here relatively smoothly.
Let an ideal $(\maW)\subseteq\CFe[S_1,\dots,S_n]=\CFe[S]$ be generated
by its Gr\"obner basis $\maW \subseteq D[\eps][S]$.
That is, we fixed a particular monomial
ordering on $\CFe[S]$, and the leading (with respect to this ordering)
terms of $\maW$ generate the ideal of leading terms of $(\maW)$. 
Following \cite[Sect.~11.3]{BPR03}, we call $\maW$ 
{\em parametrized special} if it is of the form 
$\mathcal{W}=\{b_1 S_1^{d_1}+U_1, \dots, b_n S_n^{d_n}+U_n\}$, 
where the leading terms are
$b_1S_1^{d_1}$,\dots, $b_n S_n^{d_n}$, 
and $\deg(U_i)<d_i$, $\deg_{S_j}(U_i)<d_j$ for
$1\leq i,j\leq n$.
Note that the quotient algebra $\CFe[S]/(\maW)$ has the 
natural basis $\mathrm{U}(\mathcal{W})$ of {\em monomials under the staircase},
that is, of monomials $S^\alpha=S_1^{\alpha_1}\dots S_m^{\alpha_n}$
with $\alpha_i<d_i$ for $1\leq i\leq n$.
In particular, the dimension of the quotient algebra is
$d_1\dots d_n$. In order to keep doing the arithmetic in $D[\eps]$
when reducing with respect to $\maW$ (and this is one of the purposes
of a parametrized special Gr\"obner basis), 
one works in the basis 
$\oU(\mathcal{W})=\{b^{|\alpha|} 
S^\alpha\mid S^\alpha\in\mathrm{U}(\mathcal{W})\}$, where 
$|\alpha|=\alpha_1+\dots+\alpha_n$ and 
$b$ a common multiple of $b_1$,\dots,$b_n\in D[\eps]$.

The basis $\mathcal{W}$ naturally
appears when critical points of a coordinate projection of a certain
special type of hypersurface are computed, as in \cite[Sect.~11.6]{BPR03},
in contrast to a common situation when calculation of a Gr\"obner basis 
of an ideal is computationally very costly (the latter 
can generally require
doubly exponential, in the number of variables, running time, 
cf. \cite{MR84g:20099}).
Given $\mathcal{W}$, 
one can efficiently compute the multiplication table of the
quotient algebra, see \cite[Alg.~11.22]{BPR03}. Namely,
when the degrees of the elements of $\mathcal{W}$ in
$S$ (resp. in $\eps$) are bounded by $d$ (resp.  by $\lambda$) 
it takes $(d\lambda)^{O(n\ell)}$ operations in $D$, the $\eps$-degrees
never exceed $\lambda (nd)^{O(1)}$;
when $D=\Z$, the bitsize of the data involved is bounded by
a polynomial in $n$ and $(\lambda d)^{n\ell}$ times 
the bitsizes of the elements of $\mathcal{W}$, cf. \cite[pp.~381--382]{BPR03}.
\begin{theorem}\label{thm:proj-limit-comp}
For $D$ a computable subring of $\F$, 
let $\mcB\subset D[\eps][X_1,\dots,X_\qq]$ define 
a 0-dimensional polynomial system, that is its own 
special Gr\"obner basis with the LCM of the leading terms
equal to $b_\mcB\in D[\eps]$. Let 
$$N=\dim \CFe[X]/(\mcB)\quad\text{and}\quad \CZ=Z(\mcB,\CFe^\qq).$$ 
Assume the degrees of the elements of $\mcB$
in $X$ as well as in $\eps$, and the degree of $b_\mcB$ in $\eps$, 
bounded by $d$.
Let $\PP$ be a polynomial mapping $X\mapsto (\PP_1(X),\dots,\PP_m(X))$,
with $\PP_i\in D[\eps][X]$ of degree $<d$ in $X$ and at most $d$ in 
$\eps$.\\
Then a set of at most $(m-1)N^3$ candidates for univariate representations
$u(T)\in D^{m+2}$ 
for the elements of $\lim_\eps\PP(\CZ)$ %(for each possible value of 
%certain multiplicity
%$1\leq \mu(x)\leq N$, for $x\in\lim_\eps\PP(\CZ)$) 
can be computed in $(m+N)^{O(\ell)}$ arithmetic
operations in $D$.
The degrees of the polynomials in $u(T)$ are at  
most $N$. \\
When $D=\Z$, the bitsizes of the coefficients of 
$u(T)$ and of the intermediate data
are bounded by a polynomial in $d$, $m$ and $N^\ell$ 
times the bitsize of the input data $\mcB$.
%$O(d^4n^3)(\tau_\mcB+O(\ell\log_2 nd))$,
%where $\tau_\mcB$ is a bound on the bitsizes of the coefficients
%of the elements of $\mcB$.
\end{theorem}
Note that $\mcB$ will be constructed in Theorem~\ref{thm:lim-im-samples}.
\begin{remark}
The number $(m-1)N^3$ of candidates can be reduced to $N$, 
see Remark~\ref{fno:p8} later in this section.
%
%The assumption that $\mcB$ is its own special Gr\"obner basis
%is largely technical. If one needs limits of rational images of
%{\em real} roots of a 
%0-dimensional polynomial system, the technique of the next
%section (cf. Theorem~\ref{thm:lim-im-samples}) solves the problem.
\end{remark}

The remainder of the section is devoted to the proof of
Theorem~\ref{thm:proj-limit-comp}.

\subsection{On $\PP$-separating elements}
Here we prepare the ground for the limit computations.
Denote by $\charpol{a}{T}$ the characteristic polynomial
of a linear transformation $a\in A.$
The Stickelberger's lemma \cite[Thm.~4.69]{BPR03} states in 
particular that
\beq{stickel0}
\charpol{a}{T}=\prod_{x\in\CZ} (T-a(x))^{\mu(x)}.
\eeq
where  $\mu(x)$ is the multiplicity of $x$ as a root of $\mcB$.
For a given $a\in A$, denote by $[x]\subseteq\CZ$ the equivalence
class of $x$ with respect to the equivalence relation defined
by $a$, so that $x$ is equivalent to $y$ when $a(x)=a(y)$.
Then 
\beq{stickel1}
\charpol{a}{T}=\prod_{[x]\subseteq\CZ} (T-a(x))^{\mux},\quad
\text{where}\quad \mux=\sum_{y\in [x]} \mu(y).
\eeq
Let $0\neq b\in A$ 
be an $a$-\emph{class function} on the relation $[*]$ induced by $a$, 
that is, $b(y)=b(x)$ for any $y\in [x]$.
Let $S$ be a variable and consider polynomials 
\begin{align}
\label{eq:charST}
\charpol{a+Sb}{T}&= \prod_{[x]\subseteq\CZ} (T-a(x)-Sb(x))^{\mux},\\
\label{eq:gpoldef}
\gpol{a}{b}{T}&=\frac{\partial\charpol{a+Sb}{T}}{\partial S}|_{S=0}.
\end{align}
Then 
\beq{g-def}
\gpol{a}{b}{T}=-\sum_{[x]\subseteq\CZ}b(x)\mux(T-a(x))^{\mux-1}
\prod_{[x]\neq[y]\subseteq\CZ}(T-a(y))^{\mu_{[y]}}.
\eeq
Observe that
%\beq{gder-def}
$$
\gpold{\mux-1}{a}{b}{T}=-b(x)(\mux)!
\prod_{[x]\neq[y]\subseteq\CZ}(T-a(y))^{\mu_{[y]}}+(T-a(x))h(T).
$$
%\eeq
for some polynomial $h(T)$.  Therefore
\beq{gder-atroot-def}
\gpold{\mux-1}{a}{b}{a(x)}=-b(x)(\mux)!
\prod_{[x]\neq[y]\subseteq\CZ}(a(x)-a(y))^{\mu_{[y]}}.
\eeq
In particular the following holds.
\begin{lemma}\label{lem:pi-sep-coord}
For a $\PP$-separating $a$, any $0\neq r\in\Fe$ and  $x\in\CZ$,
\beq{coord-expr}
\PP_i(x)=\frac{\gpold{\mux-1}{a}{r\PP_i}{a(x)}}
{\gpold{\mux-1}{a}{r}{a(x)}}\qquad\text{for each}\quad 1\leq i\leq m.
\eeq
\end{lemma}
\begin{proof}
As $a$ is $\PP$-separating, $\gpold{\mux-1}{a}{r}{a(x)}$ never vanishes,
and $[x]\neq [y]$ as soon as 
$\PP_i(x)\neq\PP_i(y).$
Thus $\PP_i(X)$ is a $a$-class function,
and \refeq{gder-atroot-def} holds
for $b=\PP_i$. Now \refeq{gder-atroot-def} implies the statement of the
lemma, as the terms 
$r(\mux)! \prod_{[x]\neq[y]}(a(x)-a(y))^{\mu_{[y]}}$ occur 
in both numerator and denominator of the right-hand side of \refeq{coord-expr}.
\end{proof}

%\begin{remark}
%In a similar vein one can obtain 
%$$\frac{p(x)}{h(x)}=\frac{\gpold{\mux-1}{a}{p}{a(x)}}
%{\gpold{\mux-1}{a}{h}{a(x)}}$$
%for any two $a$-class functions $p,h\in A$ with $h(x)\neq 0$, 
%and any $x\in\CZ$.
%\end{remark}

Note that $N=\sum_{x\in\CZ}\mu(x)$. 
To compute the coefficients of 
$\charpol{a+Sb}{T}$, it is convenient to write it as 
$$\charpol{a+Sb}{T}=\sum_{j=0}^N b_j(S) T^j,\quad\text{where}\quad
b_j\in\CF[S],\quad \deg{b_j}=N-j.$$
We need to be able to compute the 
trace $\Tr{f(S)}$ of a linear transformation $f(S)\in A[S]$.
By additivity of the trace, this is easy to do once a basis $\oU(\mcB)$ of $A$,
the multiplication table for $A$ in $\oU(\mcB)$, 
that is, a tensor $\lambda^{\iota}_{\alpha\omega}$ specifying
linear combinations 
$$\alpha\omega=\sum_{\iota\in\oU(\mcB)}\lambda^\iota_{\alpha\omega}\iota
\qquad\text{for}\quad \alpha,\omega\in \oU(\mcB),$$ 
and the expression of
$f(S)$ as a linear combination of the elements of $\oU(\mcB)$
with coefficients in $\CF[S]$ are known. 
Namely, %by applying
%e.g. \cite[Algorithm~11.15]{BPR03}.
$$\Tr{f(S)}=\Tr{\sum_{\omega\in\oU(\mcB)} f_\omega(S)\omega}=
\sum_{\alpha\in\oU(\mcB)}\sum_{\omega\in\oU(\mcB)}
f_{\omega}(S)\lambda^{\alpha}_{\alpha\omega},$$
and the computation can be done separately for each coefficient of
the polynomial $\Tr{f(S)}$.

It is well-known (cf. e.g. \cite[Thm.~4.69]{BPR03}) that
for $f\in A$
\beq{trace-formula0}
\Tr{f^j}=\sum_{x\in\CZ} \mu(x)f(x)^j,\qquad j\geq 0.
\eeq
It follows that
\beq{trace-formula}
\Tr{(a+Sb)^j}=\sum_{x\in\CZ} \mu(x)(a(x)+Sb(x))^j=
 \sum_{[x]\subseteq\CZ} \mux(a(x)+Sb(x))^j,\qquad j\geq 0.
\eeq
Then, the $b_j(S)$'s (that is, the \emph{elementary symmetric functions}
of the roots) can be computed knowing the 
\emph{power symmetric functions}
(also known as \emph{Newton sums}) 
$\Tr{a+Sb},\dots,\Tr{(a+Sb)^N}$ of the roots of 
$\charpol{a+Sb}{T}$.
Namely, the following holds (cf. \cite[(4.2),~(11.8)]{BPR03}).
$$\frac{\partial\charpol{a+Sb}{T}}{\partial T}=
\charpol{a+Sb}{T}\sum_{m\geq 0}\frac{\Tr{(a+Sb)^m}}{T^{m+1}}.$$
By equating the coefficients of $T^j$, for each $j$ satisfying 
$-1\leq j<N$, on the
both sides of the latter, and recalling that $\charpol{a+Sb}{T}$ is
monic in $T$, that is, $b_N(S)=1$, 
one obtains the following.
\begin{lemma}
Let $\charpol{a+Sb}{T}=\sum_{j=0}^N b_j(S) T^j$ 
be the characteristic polynomial
of a linear transformation $a+Sb\in A[S].$ Then
\beq{charpolcoefs}
b_i(S)=-\frac{1}{N-i}\sum_{j=1}^{N-i}b_{i+j}(S)\Tr{(a+Sb)^j},
\quad 0\leq i\leq N-1,\quad b_N=1
\eeq
gives a recurrence for $b_i(S)$'s, for $i=N-1,N-2,\dots,0$. \qed
\end{lemma}

\begin{remark}\label{rem:b_i}
Formulae similar to \refeq{charpolcoefs} are known since long time,
and attributed to \cite{LeVe}. An explicit expression for 
$b_i(S)$ in terms of a determinant of certain ``almost Toeplitz''
matrix with entries specified by $\Tr{(a+Sb)^j}$'s can be
found by using \cite[Ex.~I.2.8]{Mac95}. See also \cite[(2.14)]{Mac95}. 
\end{remark}

Using the latter lemma, we can construct $\charpol{a}{T}$ and
$\gpol{a}{b}{T}$ given $a,b\in A$. 
We do not need to compute $\charpol{a+Sb}{T}$ completely; namely,
only $S$-linear parts of $b_i(S)$ and $\Tr{(a+bS)^j}$ need to be computed, 
in view of \refeq{charpolcoefs} and \refeq{gpoldef}.
To avoid the necessity to handle
rational expressions arising from the term $\frac{1}{N-i}$ in 
\refeq{charpolcoefs}, compute $N!\, \charpol{a}{T}$ and
$N!\, \gpol{a}{b}{T}$ instead.

In what follows we restrict ourselves to separating elements of the
form $a(P(X))$, for $a\in D[T_1,\dots,T_m]$. 
Note that a $\PP$-separating $a$
exists and can be chosen as follows, for some 
$0\leq j\leq (m-1)\binom{N}{2}$:
\beq{well-sep-choice}
a(P(X))=a(j,P(X))=\sum_{i=1}^{m} j^{i-1}\PP_i(X).
\eeq
To see this, one proceeds as in e.g. \cite[Lemma~4.60]{BPR03}.
Let $s\neq y\in\PP(\CZ)$, and observe that the univariate polynomial
$a(Y,s)-a(Y,y)=\sum_{i=1}^m (s_i-y_i)Y^{i-1}$ is not identically 0, and has
at most $m-1$ roots. Thus
by avoiding at most $m-1$ values of $j$, one can make sure
that $a$ separates $s$ and $y$. As there are at most $\binom{N}{2}$
distinct pairs of $s$ and $y$ as above, the claim follows.

Thus we can construct univariate representations of the 
elements $s\in\PP(\CZ)$ of multiplicity $\mu(s)=\mu+1$ in the form
\begin{multline}\label{eq:univrep-proj}
u(T) = N!\, (\charpol{a}{T},\gpold{\mu}{a}{r}{T},
\gpold{\mu}{a}{r\PP_1(X)}{T},\dots,\\
\gpold{\mu}{a}{r\PP_m(X)}{T}),
\end{multline}
where $r\in D[\eps]$ is chosen so that the functions $ra(X)$, 
$r\PP_1(X)$,\dots, $r\PP_m(X)$ of $A$ are $D[\eps]$-linear
combinations of the basis elements of $A$.
The latter are chosen so that the entries of the multiplication table
of $A$ belong to $D[\eps]$, as dictated in turn by the coefficients of
the leading monomials in $\mcB$.
Taking $r$ to be the LCM $b_\mcB$ of
the coefficients of the leading monomials in $\mcB$ suffices;
the numbers $\mu(s)\leq N$ are not known {\em a priori}, thus we have
roughly $N$-fold redundancy in the output.

To summarize, we have the following.
\begin{proposition}\label{prop:gpol-comp}
Let $\mcB$, $\PP$ be as in Theorem~\ref{thm:proj-limit-comp},
$A=\CFe[X]/(\mcB)$ be of dimension $N$, and
$a\in \Z[T_1,\dots,T_m]$ 
define a $\PP$-separating element $a(\PP(X))$ given by \refeq{well-sep-choice}
with coefficients in $\Z$ of size
at most $O(\log mN)$.\\
Then a set of at most $N^{O(1)}$ univariate representations 
$u(T)\in D[\eps]^{m+2}$ 
of the form \refeq{univrep-proj},
containing 
for each 
$s\in\PP(\CZ)$, 
a representation \refeq{univrep-proj} 
with $\mu=\mu(s)-1$,
can be computed in $N^{O(\ell)}$ arithmetic operations in $D$.
The degrees of the polynomials in $u(T)$ are at
most $N$, and their coefficients are of degree 
at most $O(d^3)$ in $\eps$. \\
When $D=\Z$, the bitsize of the coefficients of 
$u(T)$ and of the intermediate data
is bounded as stated in Theorem~\ref{thm:proj-limit-comp}.
\end{proposition}
The complexity analysis is very similar to  algorithms in 
\cite[Chapter~11]{BPR03}. The most expensive part is computing
the appropriate multiplication table for $A$, see the exposition
preceding Theorem~\ref{thm:proj-limit-comp} and
 \cite[Alg.~11.22]{BPR03}), 
and this is identical
to the special case considered in [{\em loc.cit.}].
Note that $\PP_i$'s are expressed as linear combinations of the 
basis elements of $A$ and
therefore our setting for complexity analysis of computation of
$u(T)$ is essentially the same as in [{\em loc.cit.}].

\subsection{Computation of the limit}\label{subs:comli}
We proceed to computing the limits of the points given by 
univariate representations $u(T)$
of Proposition~\ref{prop:gpol-comp}.
We show that the limits of points in $\PP(\CZ)$ correspond to the 
limits $S_{<\infty}$ of
bounded roots of $\charpol{a}{T}$. Then we normalize the polynomials
in $u(T)$ by a Puiseux monomial in $\eps$ that 
makes the coefficients of $\charpol{a}{T}$ and of the rational functions
$\PP_i(T)=\frac{\gpold{\mu}{a}{bX_i}{T}}{\gpold{\mu}{a}{b}{T}}$, 
for each $1\leq i\leq m$, bounded. 
At the same time the values of the limit of the normalized
denominator $\gpold{\mu}{a}{b}{T}$ will be 
nonzero on $S_{<\infty}$. Thus $\lim_\eps\PP_i(T)$ at $S_{<\infty}$ 
can be computed by taking the limits of the coefficients of
$\PP_i(T)$, and then evaluating on the elements of $S_{\infty}$. We will give
explicit formulae for $\lim_\eps \PP_i(T)$ in terms of the appropriate
repeated derivatives of the numerator and of the denominator.

We need some further notation related to a real closed
extension $\Fz$ of a real closed field $\F$ by infinitesimals
$\zeta_1\gg\zeta_2\gg\dots\gg\zeta_\ell$, and its algebraic closure
$\CFz$. 
Let $0\neq\tau\in\CFz$. Then
$\tau$ can be written uniquely as $\tau=\zeta^{o(\tau)}(\In{\tau}+\tau')$,
with $0\neq\In{\tau}\in\CF$ and $o(\tau)\in \Q^\ell$ 
such that $\zeta^{o(\tau)}$
is the biggest, with respect to the order in $\Fz$, $\zeta$-monomial of $\tau$,
and $\tau'$ satisfying $\lim_\zeta\tau'=0$. In particular $\tau'$ is
bounded over $\F$, and $\lim_\zeta\tau=\In{\tau}$ iff $o(\tau)=0$.
Further, for $0\neq v\in\CFz^{n}$ define 
$o(v)=\max\limits_{1\leq j\leq n} o(v_j)$, with $\max$ taken 
in the sense of the ordering in $\Fz$.

Note that boundedness of $\tau=\Re\tau+i\Im\tau\in \CFz$ 
is understood here and elsewhere in this section 
in the usual sense of the  norm $\sqrt{(\Re\tau)^2+(\Im\tau)^2}$
being bounded over $\F$. As well, $\lim_\zeta\tau$
is understood purely algebraically, that is, in the appropriate order
setting to 0 the corresponding infinitesimals.
In just introduced notation, $\tau$ is bounded if and only if
either  $o(\tau)=0$, or the rightmost
nonzero entry of $o(\tau)$ is positive.  
%It can happen that $\lim_\zeta\tau\neq 0$, while $\Im\tau\neq 0$ and 
%$\lim_\zeta\Im\tau=0$, that is, taking limit makes a point real.

Let $f(T)=\sum_{j=0}^m c_j T^j\in\F(\zeta)[T]$.
Then $o(f)$ is defined to be such that $\zeta^{o(f)}$ is minimal,
with respect to the order in $\Fz$, monomial making
$\zeta^{-o(f)}c_j$ for $0\leq j\leq m$ bounded over $\F$.
In fact $o(f)=o(c_j)$ for some $j$. Define 
\beq{hatdef}
\hat{f}(T)=\lim_{\zeta}\zeta^{-o(f)}f(T),
\eeq
where the limit is taken coefficient-wise.
For $F(T)\in \F(\zeta)[T]^\qq$, we denote 
\beq{defoF}
o(F)=\max_{1\leq i\leq k} o(F_i).
\eeq
%For a multiset $\Sigma$ of elements of $\CFz^\qq$ and 
%$y\in\lim_\zeta\Sigma$, define 
%\beq{liminv}
%\lim_\zeta^{-1}y=\{x\in\Sigma\mid y=\lim_\zeta x\},\qquad
%\mu(y)=\sum_{x\in \lim_\zeta^{-1}y}\mu(x).
%\eeq
The following statement is an extension of
Lemma~11.37 from \cite{BPR03} adjusted to the non-multiplicity-free
situation, and will be used repeatedly.
\begin{lemma}\label{lem:1137} 
Let $f(T)\in\F(\zeta)[T]$ be monic. Denoting 
$Z_f=Z(f(T),\CFz)$, one has 
$$o(f)=\sum_{\substack{\tau\in Z_f,\\\text{unbounded}}}
\mu(\tau)o(\tau),\qquad Z(\hat{f}(T),\CF)=\lim_\zeta Z_f.$$
Let $y\in Z(\hat{f}(T),\CF)$.  Then 
$\mu(y)=\sum\limits_{{\substack{\tau\in Z_f,\\\lim_\zeta\tau=y}}} \mu(\tau)$ 
equals the multiplicity of $y$ as a root of $\hat{f}$. Here
the summands $\mu(\tau)$ denote multiplicities of  roots $\tau$ of $f$.
\end{lemma}
\begin{proof} (Sketch.)
The coefficients $f_i$ of $f=\sum_{i=0}^d f_i T^i$, where $f_d=1$, 
are elementary symmetric functions of $x\in Z_f$ 
taken with multiplicities. Let $\Sigma$ denote the multiset
of roots of $f$. Then
\beq{f_iinroots}
f_{d-i}=\sum_{\substack{\Theta\subseteq\Sigma\\|\Theta|=i}}\ 
\prod_{\tau\in\Theta}\tau,
\eeq
implying
$$o(f_{d-i})\leq\max_{\substack{\Theta\subseteq\Sigma\\
|\Theta|=i}}\ \sum_{\tau\in\Theta}o(\tau),$$
where the inequality might be struct due to 
a possible cancellation of higher order
terms in the sum \refeq{f_iinroots}.
When one of the multisets $\Theta$ equals $\Upsilon$, 
the sub-multiset of unbounded roots of $f$, this inequality turns into 
equality $\overline{o}=o(f_{d-|\Upsilon|})=\sum_{\tau\in\Upsilon}o(\tau)$, 
as the order of the remaining summands in \refeq{f_iinroots}
is strictly less than $\overline{o}$.
As $\overline{o}\geq\sum_{\tau\in\Theta}o(\tau)$
for any $\Theta\subseteq\Sigma$, and we obtain 
$o(f)=\overline{o}$.
Therefore
\begin{align}
\label{eq:hatfT1}
\hat{f}(T)&=\lim_\zeta\eps^{-o(f)}f(T)=
\lim_\zeta \prod_{\tau\in\Sigma-\Upsilon}(T-\tau)
 \prod_{\tau\in\Upsilon}\eps^{-o(\tau)}(T-\tau)=\\
 &=\prod_{\tau\in\Upsilon}(-\In{\tau})
\prod_{\tau\in\Sigma-\Upsilon}(T-\lim_\zeta\tau),
\label{eq:hatfT}
\end{align}
and the first part of the lemma follows.
The second part follows from \refeq{hatfT}.
\end{proof}

A $\PP$-separating element $a$ will be called \emph{well-$\PP$-separating}
(with respect to $\lim_\eps$) if the following conditions hold:
\begin{enumerate}
\item for any $s,y\in\PP(\CZ)$ such that 
$\lim_\eps s\neq \lim_\eps y$ one has 
$a(\lim_\eps s)\neq a(\lim_\eps y)$;\label{cond-well-sep1}
\item $o(\PP(u))=o(a(\PP(u)))$ for any $u\in\CZ$.\label{cond-well-sep2}
%\item $o(\charpol{a}{T})\geq o(\gpol{a}{X_i}{T}$ for all $1\leq i\leq m.$
%\item $o(\charpol{a}{T})\geq o(\gpol{a}{1}{T}) \geq o(\gpol{a}{b}{T})$ 
%for any
%$a$-class function $0\neq b\in A$ given by a  element of $\F[\PP(X)]$
%of degree not exceeding the degree of $a$.
%\label{cond-well-sep3}
\end{enumerate}
In particular, \ref{cond-well-sep2} implies that
if $s\in\PP(\CZ)$ is unbounded over $\F$ then 
$a(s)$ is also unbounded over $\F$.
%Note that $o(\charpol{a}{T})\geq o(\gpol{a}{1}{T})$, due to  
%\refeq{gpoldef}. Thus the condition \ref{cond-well-sep3} allows one,
%in view of \refeq{hatdef},
%to compute the limit of $b$ at the elements of $\CZ$  after
%multiplying (appropriate derivatives of) 
%$\gpol{a}{1}{T}$ and $\gpol{a}{b}{T})$ by 
%$\eps^{-o(\charpol{a}{T})}$.

%From this point on we restrict ourselves to the case of \emph{linear}
%$a$-class functions $b=sum_j \beta_j \PP_j(X)$, with $\beta\in\Z^m$.
\begin{lemma}\label{lem:bx-lim}
Let $a,b\in D[T_1,\dots,T_m]$ be linear. 
Let $a$ be well-$\PP$-separating.
Then for any $s\in\lim_\eps\PP(\CZ)$ of multiplicity $\mus$
$$b(s)=\frac{\gpolhatd{\mus-1}{a}{b}{a(s)}}
{\gpolhatd{\mus-1}{a}{1}{a(s)}}.\qquad\text{In particular,}\quad
s_j=\frac{\gpolhatd{\mus-1}{a}{\PP_j}{a(s)}}
{\gpolhatd{\mus-1}{a}{1}{a(s)}},$$
for any $1\leq j\leq m$.
\end{lemma}
\begin{proof}
As $a$ is well-$\PP$-separating, $b$ is an $a$-class function on 
$\PP(\CZ)$ as well as on $\lim_\eps\PP(\CZ)$.
Note that $b$ satisfies
$\lim_\eps b(\PP(x))=b(\lim_\eps \PP(x))$, for any $x\in\CZ$ for which
$\lim_\eps \PP(x)$ is defined. 
We shall express  $\lim_\eps b(\PP(X))$ as a univariate rational function, that
gives  $\lim_\eps b(\PP(y))$ when evaluated at $a(\PP(y))$.  With $a=a(\PP(X))$
and $b=b(\PP(X))$, denote
$$\eta(a,b,S,T)=\lim_\eps\eps^{-o(\charpol{a}{T})}\charpol{a+Sb}{T}.$$
\newcommand{\CZb}{\CZ_{<\infty}}
Denoting by $\CZb$ the set of $x\in\CZ$ such that $a(P(x))$ is bounded, 
using \refeq{charST} and Lemma~\ref{lem:1137} (in particular \refeq{hatfT})
one obtains  
\begin{align*}
\eta(a,b,S,T)&=\lim_\eps\eps^{-o(\charpol{a}{T})}
\prod_{[x]\subseteq\CZ} (T-a(\PP(x))-Sb(\PP(x)))^{\mux}=\\
&=\prod_{[x]\subseteq\CZb}\lim_\eps (T-a(\PP(x))-Sb(\PP(x)))^{\mux}\times\\
&\times\prod_{[x]\subseteq\CZ-\CZb}\lim_\eps
\left(\eps^{-o(a(P(x)))}(T-a(\PP(x))-Sb(\PP(x)))\right)^{\mux}=\\
&=G(S)\prod_{[y]\subseteq
\lim_\eps\PP(\CZ)}(T-a(y)-Sb(y))^{\muy}, \quad\text{where}\\
G(S)&=\prod_{[x]\subseteq\CZ-\CZb}
(-\In{a(\PP(x))}-Sb(\lim_\eps \eps^{-o(P(x))}\PP(x)))^{\mux}\in\CF[S],
\end{align*}
as $o(a(P(x)))=o(P(x))$. Moreover, 
$G(0)\neq 0$, as $\In{a(\PP(x))}\neq 0$ on
unbounded $x\in\PP(\CZ)$.  

In view of \refeq{hatfT} one obtains in particular 
\beq{charpolhat}
\hat{\chi}(a,T)=C\prod_{[y]\subseteq\lim_\eps\PP(\CZ)}(T-a(y))^\muy,
\quad\text{with}\quad 0\neq C\in\CF,
\eeq
where $\muy$ denotes the multiplicity of $a(y)$ as a root
of $\hat{\chi}(a,T)$.

Next, we compute, for $\gpol{a}{b}{T}$ defined by \refeq{gpoldef}, 
$\lim_\eps\eps^{-o(\charpol{a}{T})}\gpol{a}{b}{T}$. 
We see that it equals to $\gpolhat{a}{b}{T}$ 
(if it would not be the case, it had to vanish identically, as
$o(\charpol{a}{T})\geq o(\gpol{a}{b}{T})$), 
as defined in \refeq{hatdef}, so we get
\begin{multline*}
\gpolhat{a}{b}{T}=\frac{\partial}{\partial S}\eta(a,b,S,T)|_{S=0}=
-G'(0)\prod_{[y]\subseteq\lim_\eps\PP(\CZ)}(T-a(y))^{\muy}-\\
-G(0)\sum_{[y]\subseteq\lim_\eps\PP(\CZ)}\muy b(y)(T-a(y))^{\muy-1}
\prod_{[y]\neq [s]\subseteq\lim_\eps\PP(\CZ)}(T-a(s))^{\mus}.
\end{multline*}
Then, for any $y\in\lim_\eps\PP(\CZ)$,
$$\gpolhatd{\muy-1}{a}{b}{T}=-G(0)b(y)(\muy)!
\prod_{\substack{[s]\subseteq\lim_\eps\PP(\CZ)\\ [y]\neq [s]}}
     (T-a(s))^{\mus}+(T-a(y))H(T),$$
for $H(T)\in\CF[T]$. Hence
$$\gpolhatd{\muy-1}{a}{b}{a(y)}=-G(0)b(y)(\muy)!
\prod_{[y]\neq [s]\subseteq\lim_\eps\PP(\CZ)}(a(y)-a(s))^{\mus}$$
and we obtain,  in view of \refeq{charpolhat},
the statement of the lemma. 
\end{proof}

Let us show that $a$ can be taken to be 
$a=a(j,\PP(X))$ for a certain $j$ as in \refeq{well-sep-choice}.
The only difference with the argument above is that we have to 
avoid more ``wrong'' values of $j$.
Let $x,y\in\PP(\CZ)$ be such that $\lim_\eps x$ and $\lim_\eps y$ exist
and are not equal. Then the polynomials
$a(Y,\lim_\eps x)-a(Y,\lim_\eps y)$ and 
$a(Y,x)-a(Y,y)$ are not identically 0 and each of them has at
most $m-1$ roots.
Thus by avoiding at most $2(m-1)$ values of $j$, one can make sure
that $a$ separates $\lim_\eps x$ and $\lim_\eps y$, as well as $x$ and $y$.

To ensure the remaining condition in the definition of well-$\PP$-separating
element, consider $W=\{\lim_\eps \eps^{-o(s)}s\mid 0\neq s\in\PP(\CZ)\}$.
Choose $j$ such that $a(j,w)\neq 0$ for any $w\in W$.
Such a choice is always possible: $a(Y,w)\in\F[Y]$ has at most 
$m-1$ roots; thus avoiding $(m-1)|W|$ values of $j$ achieves the required.
Then $o({a(s)})=o(s)$ for all $s\in\PP(\CZ)$, 
implying condition \ref{cond-well-sep2} of the definition. 

%Note that $\gpold{\mu-1}{a}{b}{T}$ is defined via the $\mu$-th partial 
%derivative of $\charpol{a+Sb}{T}$. The boundedness of the coefficients
%of 
%$$\rho(a,b,S,T)=\eps^{-o(\charpol{a}{T})}\charpol{a+Sb}{T}$$ 
%will imply boundedness of the coefficients of 
%$\eps^{-o(\charpol{a}{T})}\gpold{\mu-1}{a}{b}{T}$, that is, the
%remaining condition \ref{cond-well-sep3} of the definition.
%Denoting by $\CZb$ the set of $x\in\CZ$ such that $a(x)$ is bounded, 
%one has,
%by the first part of Lemma~\ref{lem:1137}, that
%\begin{multline*}
%\rho(a,b,S,T)=
%\prod_{[x]\subseteq\CZb}(T-a(x)-Sb(x))^{\mux}\times\\
%\times\prod_{[x]\subseteq\CZ-\CZb}\left(\eps^{-o(x)}
%  (T-a(x)-Sb(x))\right)^{\mux}
%\end{multline*}
%has bounded coefficients. Indeed, for unbounded $x\in\CZ$ the 
%multiplicative contributions
%in the coefficients are of the form $\eps^{-o(x)}a(x)$ and 
%$\eps^{-o(x)}b(x)$,  that are bounded, as the choice of
%$j$ implied $o(a(x))=o(x)$, and as $o(a(x))\geq o(b(x))$ by assumptions on 
%$b$.

As there are at most $\binom{N}{2}$
distinct pairs of $s$ and $y$ as above, and since $|W|\leq N$, we obtain
\begin{lemma}\label{lem:wel-pi-sep-elt}
There exists an integer 
$0\leq j\leq (m-1)N^2$ such that $a(\PP(X))=a(j,\PP(X))$ 
is well-$\PP$-separating. \qed
%and $o(a(x))=o(x)$ for any $x\in\PP(\CZ)$. \qed
\end{lemma}

Combining Lemmas~\ref{lem:bx-lim} and \ref{lem:wel-pi-sep-elt}
gives for an appropriate $a$ at most $O(\sqrt{N})$ candidates for
univariate representations 
$u(T)$ for
the points in $\lim_\eps\PP(\CZ)$, as $\charpol{a}{T}$ has at most 
$O(\sqrt{N})$ different root multiplicities.
We outline now how $u(T)$ are actually computed. 
Let $b=\PP_i$ for some $1\leq i\leq m$. We loop through $1\leq j\leq (m-1)N^2$
in order to be sure to 
find an appropriate well-$\PP$-separating $a(\PP(X))=a(j,\PP(X))$.
This means that we will return candidate representations for each such $j$.

A further technical point
is that we operate in the ring $D[\eps]$ rather than in a field.
We utilize the idea of \cite[Remark~11.44]{BPR03} for $r=b_\mcB$: 
\beq{rem1144}\charpol{ra+Srb}{rT}= r^N\charpol{a+Sb}{T}\eeq
and proceed similarly to the procedure of \cite[Alg.~11.45]{BPR03}.
\begin{remark}\label{fno:p8}
Compared to [{\em loc.cit.}], our 
simplification is that we do not try to filter out ``wrong'' 
$a$; such a check would require finding $|\PP(\CZ)|$ to be able to verify that
$a$ is $\PP$-separating, and then proceeding similarly to remarks on
\cite[p.398]{BPR03}. 
To obtain ``good'' $a$, one would first select
$a$'s with the biggest degree of $\charpol{a}{T}$; among the latter
select $a$'s that are $\PP$-separating, by choosing $a$'s with minimal
degree of $\mathrm{gcd}(\charpol{a}{T},\frac{d\charpol{a}{T}}{dT})$.
%Among the latter, select $a$'s with minimal degree of $\hat{\chi}(a,T)$,
%thus ensuring that $a$ maps unbounded elements of $\PP(\CZ)$ to
%unbounded elements of $\Fe$. 
To ensure that $a$ is well-$\PP$-separating,
one would select $a$'s with minimal degree of
$\mathrm{gcd}(\hat{\chi}(a,T),\frac{d\hat{\chi}(a,T)}{dT})$.
Finally, to make sure that $o(a(x))=o(x)$ for any $x\in\PP(\CZ)$,
one would check that $o(\charpol{a}{T})\geq o(\gpol{a}{\PP_i}{T})$ for all 
$1\leq i\leq m$.

Anyhow, this shortcut does not worsen the asymptotic running time. 
\end{remark}

Thus we compute $\charpol{ra+Srb}{T}$, as already 
outlined in the first part of this section, and then operate with  
$\eps^{-o(\charpol{ra}{rT})}\charpol{ra+Srb}{rT}$ to obtain the remaining
data for the limit computation. We have 
$$\hat{\chi}(ra+Srb,rT)=
 \lim_\eps\eps^{-o(\charpol{ra}{rT})}\charpol{ra+Srb}{rT}.$$
Using the latter to compute $\gpolhat{ra}{rb}{bT}$ and 
$\gpolhat{ra}{r}{bT}$ as in the proof of Lemma~\ref{lem:bx-lim},
by \refeq{rem1144} we obtain 
$b(x)=\frac{\gpolhatd{\mu}{ra}{rb}{ra(x)}}
 {\gpolhatd{\mu}{ra}{r}{ra(x)}}$ for
each $x\in\lim_\eps\PP(\CZ)$ of multiplicity $\mu+1$, as required.
For each $j$ we loop through all the possible values of $\mu$.
Thus in total we will have no more than 
$(m-1)N^3$ univariate representations, 
as stated in the theorem.

The complexity analysis for a given $j$ 
runs parallel to the analysis given in \cite[Sect.~11.5]{BPR03} 
(see also a remark following 
Proposition~\ref{prop:gpol-comp} above) and is omitted.
This completes the proof of 
Theorem~\ref{thm:proj-limit-comp}.

\section{Limits of solution images: dimension $>0$}\label{sect:limimage}
In Section~\ref{sect:limimagedim0}
we described a procedure to compute limits of images
of finite algebraic sets under polynomial mappings. 
To complete the proof of Theorem~\ref{thm:imlim},
we proceed to computing limits,
with respect to $\eps\rightarrow 0$, 
of rational images of samples of
connected components of real algebraic sets, by reducing to the 0-dimensional
case.

For $F_0\in D[\eps][Y_1,\dots,Y_{\qq-1}]$, 
let  $Z_0=Z(F_0(Y),\Fe^{\qq-1})$, and let $\Psi$ be as in (\ref{def:Psi}).
We want to find  points in each connected component of
$\lim_\eps\Psi(Z_0)\subseteq \F^m$. 
\begin{theorem}\label{thm:lim-im-samples}
Let $F_0$ and $\Psi$ be as above, with the $Y$-degree of $\F_0$ at most $d$
and the $Y$-degrees of $\Omega_i$ and $\Lambda$ less than $d-1$, and
their $\eps$-degrees at most $d$.
Then one can construct a $(q+2)$-variate 
0-dimensional polynomial system $\mathcal{B}$ over the real closed
transcendental extension $\F\langle\eps_1,\dots,\eps_{\ell+2}\rangle$
of $\Fe$ by two infinitesimals 
$0<\eps_{\ell+2}\ll\eps_{\ell+1}\ll\eps_\ell$, 
with zero set $Z(\mathcal{B})$ such that
$$\mathcal{L}=\lim_{\eps_1,\dots,\eps_{\ell+2}} \Psi(\pi(Z(\mathcal{B}))),$$
where $\pi$ is the orthogonal projection to $Y_1,\dots,Y_{q-1}$, 
intersects each connected component of 
$\lim_{\eps_1,\dots,\eps_\ell}\Psi(Z_0)$.\\
The system $\mathcal{B}$ can be constructed in 
$(m+d)^{O(\qq\ell)}$ operations over $D$, and it 
satisfies the properties required by
Theorem~\ref{thm:proj-limit-comp}; the set 
$\mathcal{L}$ can be computed using the procedure of 
Theorem~\ref{thm:proj-limit-comp}.
\end{theorem}

In the proof we will denote
$\mu=\eps_{\ell+1}$ and $\zeta=\eps_{\ell+2}$.

First of all, let us reduce the setting to the case when $\Lambda=1$.
Assume that $F_0(Y)\geq 0$, otherwise replace $F_0$ with $F_0^2$.
Consider
$$F(Y_1,\dots,Y_\qq)=F_0(Y)+(1-Y_\qq\Lambda(Y))^2,\qquad Z=Z(F(Y),\Fe^\qq).$$
Then $\Psi(Z_0)=\PP(Z)$, where $\PP\in D[\eps][Y_1,\dots,Y_\qq]^m$
is given by 
$$(Y_1,\dots,Y_\qq)\mapsto (Y_\qq\Omega_1(Y),\dots,Y_\qq\Omega_m(Y)).$$
Thus from now on we consider the problem of computing
representatives of connected components of
the set $\lim_\eps\PP(Z)\subseteq \F^m$.
Note that the map $\PP$ is as required by 
Theorem~\ref{thm:proj-limit-comp}.

The idea is to look for points of minimal norm in the connected
components of $\PP(Z)$.
We will repeatedly use the following statement.
\begin{proposition}\label{prop:11.56}
(\cite[Prop.~11.56]{BPR03})
Let $S\subseteq\Ke^n$ be a semialgebraic set, for $\Ke$ a real closed
extension of a real closed field $\K$ by an infinitesimal $\eps$.
Then $\lim_{\eps} S\subseteq\K^n$ is a closed semialgebraic set.
If $S$ is in addition connected and bounded over $\K$ then 
$\lim_{\eps} S$ is connected. \qed
\end{proposition}

Introduce another variable $Y_0$ that will help us in this task;
we will write down, after necessary preparations,
equations that specify the critical points
of the projection $Y\mapsto Y_0$. Replace $F$ by
$$F'(Y_0,\dots,Y_\qq)=F(Y)+(Y_0-\sum_{i=1}^m\PP_i(Y)^2)^2.$$
We extend $\PP$ to the zero set $Z'$ of $F'$ by ``ignoring'' $Y_0$.
Obviously $\PP(Z)=\PP(Z')$.

\newcommand{\Fem}{\F\langle\eps,\mu\rangle}
Next, we need to make the set $Z'$ bounded, in the standard way of e.g.
\cite[Chapt.~11]{BPR03}. Introduce an infinitesimal $0<\mu\ll\eps_\ell$,
a variable $Y_{\qq+1}$ and define
\newcommand{\ZmuB}{Z_\mu^{<\infty}}
$$ %\beq{defF1}
F_\mu(Y_0,\dots,Y_{\qq+1})=F'(Y)+(1-\mu^2\sum_{j=0}^{\qq+1}Y_j^2)^2,\qquad
\ZmuB=Z(F_\mu,\Fem^{\qq+2}).
$$ %\eeq
Again, extend $\PP$ onto $\ZmuB$ by ``ignoring'' $Y_{\qq+1}$.  Define 
$$ %\beq{defZmu}
Z_\mu=Z(F',\Fem^{\qq+1})
$$ %\eeq
\begin{lemma}\label{lem:limZmu}
The following holds:
\begin{enumerate}
\item
$\PP(\ZmuB)$ is closed;\label{limZmu-1}
\item 
$\PP(\ZmuB)\subseteq \PP(Z_\mu)$;
\item
$\lim_{\mu}\PP(\ZmuB)$ equals the closure of $\PP(Z)$.\label{limZmu-3}
\end{enumerate}
\end{lemma}
\begin{proof}
The first part of the lemma follows from \cite[Thm.~3.20]{BPR03}, as
$\PP(\ZmuB)$ is the image of a closed and
bounded semialgebraic set under a continuous semialgebraic 
function.

The second part is straightforward, by observing
that the projection of $\ZmuB$ on the first $\qq+1$ coordinates
is a subset of $Z_\mu$.

Then, $\lim_{\mu}\PP(\ZmuB)$ is closed by Proposition~\ref{prop:11.56}. 
Next, $\lim_{\mu}\PP(\ZmuB)\supseteq \PP(Z)$.
Indeed, if $u=\PP(y)$ with $y\in Z$ then there exists 
$y_{\qq+1}\in\Fem$ such that 
$\mu^2\|y_\mu\|^2=1$, 
where we denoted $y_\mu=(y_0,y_1,\dots,y_{\qq+1})\in\ZmuB$, 
implying $F_\mu(y_\mu)=0$ and $u=\PP(y_\mu)\in\PP(\ZmuB)$.

Finally, the inclusion of 
$\lim_{\mu}\PP(\ZmuB)$ in the closure of $\PP(Z)$
follows from the second part.
\end{proof}

The next step is to introduce a deformation, same as in 
\cite[Sect.~11.6]{BPR03}, that ensures smoothness.  
As in \cite[(11.13)-(11.14)]{BPR03}, introduce an infinitesimal
$0<\zeta\ll\mu$, and set
\newcommand{\Def}{\mathrm{Def}}
\newcommand{\bd}{\overline{d}}
\beq{our1114}
F_\zeta(Y)=\zeta\left(\mu^{\bd}\left(Y_0^{\bd}+
\sum_{j=1}^{\qq+1}(Y_j^{\bd}+Y_j^2)\right) -2\qq-3\right)+
(1-\zeta)F_\mu(Y).
\eeq
Note that the difference between \cite[(11.13)-(11.14)]{BPR03}
and \refeq{our1114} is purely notational: variables and infinitesimals
have different
names, and they range slightly differently (i.e. there are $\qq+2$ variables
$Y$ instead of $\qq$ variables $X$). Here $\bd$ is the minimal
positive even number strictly bigger than the degree of $F_\mu$; 
this simplification of \cite[(11.13)-(11.14)]{BPR03}
is explained in \cite[Rem.~11.49]{BPR03}.

By \cite[11.50-11.51]{BPR03}, the set 
$Z_\zeta=Z(F_\zeta,\F\langle\eps,\mu,\zeta\rangle^{\qq+2})$ 
is a nonsingular algebraic
hypersurface contained in the ball of radius $\frac{1}{\mu}$, and
such that $\lim_\zeta Z_\zeta=\ZmuB$.
In particular $\lim_\zeta u$ exists for any $u\in Z_\zeta$.

Consider the set $K_\zeta$ of 
the critical points of the projection $Y\mapsto Y_0$ on $Z_\zeta$.
They satisfy the system of equations
\beq{crptsF2}
0=F_\zeta(Y)=\frac{\partial F_\zeta}{\partial Y_1}=\dots=
\frac{\partial F_\zeta}{\partial Y_{\qq+1}}.
\eeq
The zero set $\overline{K}_\zeta\supseteq K_\zeta$ 
of \refeq{crptsF2} in the algebraic closure of $\F\langle\eps,\mu,\zeta\rangle$
is analysed in detail in \cite[Sect.~11.6]{BPR03}.
In particular, $|\overline{K}_\zeta|\leq d^{O(\qq)}$
(cf. \cite[Prop.~11.57]{BPR03}).
Note that $F_\zeta(Y)$ can be reduced (cf. \cite[Notation~11.59]{BPR03})
using the relations
$\frac{\partial F_\zeta}{\partial Y_j}=0$, $j\geq 1$ so that the
degree in $Y_j$ of the result is strictly less than 
$\bd$. The result of these reductions, together with the
$\frac{\partial F_\zeta}{\partial Y_j}$, $j\geq 1$, forms
the polynomial system $\mathcal{B}$ of Theorem~\ref{thm:lim-im-samples}
 we are proving.

To complete the proof that $\mathcal{L}$ is as claimed, 
it remains to prove the following lemma.
It will make sure that we recover in $\mathcal{L}$ 
a representative of each  component of $\lim_\eps\PP(Z)$.
\begin{lemma}
The set $\lim_{\zeta,\mu}P(K_\zeta)$ intersects
each connected component of $\PP(Z)$ in a point  of minimal norm.
\end{lemma}
\begin{proof}
By construction, $\PP(K_\zeta)$ contains the local minima
of $Y\mapsto Y_0$ on $\PP(Z_\zeta)$.

By \cite[11.55]{BPR03}, the set $K_\mu=\lim_\zeta K_\zeta$
(the set of {\em pseudo-critical points})
intersects each connected component of $\ZmuB$.

As $P$ does not depend upon $\zeta$, we have that
$$\lim_\zeta \PP(Z_\zeta)=\PP(\lim_\zeta Z_\zeta)=
\PP(\ZmuB).$$
Due to closedness and boundedness over $\Fem$ of $\PP(\ZmuB)$ 
(cf. Lemma~\ref{lem:limZmu}), %the function
$Y\mapsto Y_0$ reaches its minimum %$y_0(C_\mu)$ 
on each connected component 
$C_\mu\subseteq\PP(\ZmuB)$. Note that by 
Proposition~\ref{prop:11.56} we have 
$C_\mu=\lim_\zeta C_\zeta$, for a connected
$C_\zeta\subseteq\PP(Z_\zeta)$.
Consider $\PP^{-1}(C_\mu)\subseteq \ZmuB$.
This set is a union of connected components of $\ZmuB$, and
$K_\mu$ intersects each of them; among these intersection points
there are ones with minimum value of $Y_0$.
Therefore $\PP(K_\mu)$ contains representatives of the local minima 
(that need not be singletons any more) of $Y\mapsto Y_0$ on $\PP(Z_\mu)$.

As $\PP(\ZmuB)$ is the intersection of $\PP(Z_\mu)$ with $\PP(B_{1/\mu})$, 
where $B_{1/\mu}$ denotes the ball
of radius $\frac{1}{\mu}$ and centre at the origin, 
each connected component $C$ of $\PP(Z_\mu)$ that intersects $\PP(B_{1/\mu})$
will contain a connected component $C_\mu$ of $\PP(\ZmuB)$.

We assume now, without loss of generality, that
$C$ contains a point of $\Fe$-finite norm. 
Consider the intersection $C_r$ of $C$ and the (smaller) set $\PP(B_{r})$,
where $1+\min_{u\in C}\|u\|<r\in\Fe$.
This is a bounded over $\Fe$ semialgebraic set, with finitely many connected
components $C_r^i$.
By Proposition~\ref{prop:11.56} the set 
$\lim_\mu C_r^i$  is connected, for any $i$.
As $C_r^i\subseteq C_\mu$, and as
$Y_0$ has a local minimum on $C_r^i$,
each $C_r^i\cap \PP(K_\mu)\neq\emptyset$. Thus $\lim_\mu\PP(K_\mu)$
intersects each connected component of $\PP(Z)$ in a point
of minimal norm, as required.
\end{proof}

A straightforward count of the number of operations needed to
construct $\mathcal{B}$ completes the proof of 
Theorem~\ref{thm:lim-im-samples}.

To complete the proof of Theorem~\ref{thm:imlim}, 
apply Theorem~\ref{thm:lim-im-samples} (with $\ell=2$) to $Z_0:=Z_F$
and $F_0:=F$. 
We obtain a 0-dimensional system $\mathcal{B}$ of 
equations in the real closed extension of $\Fe$ by two extra infinitesimals
$\mu=\eps_3$ and $\zeta=\eps_4$, 
with the set of solutions $\overline{K}_\zeta$ such that
$\lim_{\mu,\zeta}\Psi(\pi(\overline{K}_\zeta))$ 
intersects each connected component
of $\Psi(Z_0)$ in a point of minimal norm.
Hence applying to $\mathcal{B}$ the procedure of 
Theorem~\ref{thm:proj-limit-comp} 
will produce a set
$\lim_{\eps,\mu,\zeta}\Psi(\pi(\overline{K}_\zeta))$ intersecting each
connected component of $\lim_\eps\Psi(Z_F)$ 
in required running time, as $N\leq d^{\qq}$. \qed

\section{Pieces of extrema of a bounded level set over a quadratic map}
\label{sect:pieces}
\newcommand{\overl}[1]{\hat{{#1}}}
Let the quadratic map $Q : \K^n\rightarrow\K^k$ be given by 
\begin{equation}\label{eq:Qdef}
\begin{split}
Q_j  : X & \mapsto \frac{1}{2}X^T H_j X + b_j^T X+ c_j,\qquad
1\leq j\leq k\\
& c_j\in D,\quad b_j\in D^n,\quad H_j=H_j^T\in D^{n\times n}.
\end{split}
\end{equation}

Let $\zeta\in\K$.  We consider the $\zeta$-level set $V=Z(p(Q(X))-\zeta,\K^n)$ 
of a polynomial $p\in D[Y_1,\dots,Y_k]$ of degree $d$ 
over a quadratic map $Q : \K^n\rightarrow\K^k$ given by
(\ref{eq:Qdef}).

We assume that $D$ is a computable subring of a real closed field $\K$
and $V$ is bounded over $\K$.
The latter assumption is in fact technical. It can (and will) be ensured
in Section~\ref{sect:mainproof} by introducing an extra infinitesimal $\eps_0$
and two extra variables, to make $V$ the image of an algebraic set
on the sphere of radius $1/\eps_0$ under projection onto
the first $n$ coordinates, and subsequently removing $\eps_0$
by restricted elimination.

As well, we assume that $\zeta$ is a \emph{regular value} of 
$p(Q(X))$ and of $p(Y)$.

Let $V_c\subseteq V$ denote the set of critical points of the 
projection map $X\mapsto X_1$ from $V$ to $\K$.
Then, %as $\frac{\partial p(Q(X))}{\partial X_1}(x)\neq 0$ for
%any $x\in V_c$, 
as $\zeta$ is not a critical value of $p(Q(X))$,
$V_c$ is an algebraic set defined by
\begin{align}
\label{eq:defV}
p(Q(X))-\zeta&=0,\\
\label{eq:defVc}
\frac{\partial p(Q(X))}{\partial X_2}&= \dots
=\frac{\partial p(Q(X))}{\partial X_n}=0.
\end{align}
Due to the assumption that $V$ is bounded, 
$V_c$ intersects nontrivially
each connected component of $V,$ see \cite{GV88}, \cite[Prop.~7.6]{BPR03}. 
Thus a set $S_c$ of representatives
of each connected component of $V_c$ will also intersect each
component of  $V.$
A useful property of $S_c$ is that it will contain points in $V$
with minimal value of $X_1.$

Define
$$p_j(Y)=\frac{\partial p(Y)}{\partial Y_j},\quad\text{for}
\quad 1\leq j\leq k.$$
The set of $m\times n$ matrices $A_1,\dots,A_k$ over $\K$,
$m\leq n$,
is said to be in $r$-\emph{general position} with respect to
$V$ if
\beq{genpos}
\rank{\sum_{i=1}^k t_i A_i}\geq r\quad\text{for all}\quad
t\in (p_1(Q(V)),\dots,p_k(Q(V)))\subseteq\K^k.
\eeq
Note that $t$ is never $0$ here, as $\zeta$ is not a critical
value of $p(Y)$.
This is a weaker notion of general position than the one
used in \cite{Bar93}, where $t$ was allowed to range over
$\CK^k$, instead of  
belonging to the image of $V$ under the
polynomial mapping $X\mapsto (p_1(Q(X)),\dots,p_k(Q(X)))$. 

Note that all the nonzero $\CK$-linear  combinations
of \emph{almost} any $k$-set of  $m\times n$, $m\leq n,$ matrices
over $\K$ have rank at least $m+1-k.$

Below we assume that the matrices $\overl{H}_j$'s,
that are obtained from $H_j$'s by removing the first row, 
are in $r$-general position with respect to $V.$
For $U\subseteq\{1,\dots,m\}$ 
we denote  $\overline{U}=\{1,\dots,m\}-U.$ 
For $U\subseteq\{1,\dots,m\}$ and $W\subseteq\{1,\dots,n\}$
we denote by $A_{UW}$ the submatrix of $A$ obtained by
removing all the rows in $\overline{U}$  and all
the columns in $\overline{W}.$

To compute $S_c$ in less than exponential in $n$ time,
or even just to give  an upper bound on its size of order
less than exponential in $n$, standard methods such as
one in \cite{GV88}, \cite[Chapter~11]{BPR03} that treat
the system \refeq{defV}-\refeq{defVc} directly do not suffice.

Indeed, in these methods the number of variables, $n$ in this
case, appears unavoidably in the exponent of the bounds.
In contrast, we are able to get better results for $r$ being close to $n$.
We cover $V_c$ by (at most 
$n^{O(n-r)}$) semialgebraic sets we call \emph{pieces}, 
each of them isomorphic to a semialgebraic set lying in $\K^t$ with  
$t\leq k+n-r.$
(Here covering means simply that the union of the pieces is $V_c$; 
they in general intersect, and can even be equal one to another.)
Each of the latter is 
defined by at most $O(dn^2)$ polynomials of degree at most 
$O(nd)$.  To them, one can apply the standard technique,
see e.g. \cite{BPR03}, to bound the number of its
connected components, so there will be at most $(nd)^{O(k+n-r)}$
of them, and to find their (perhaps non-unique) 
representatives. However, for the latter,
for a technical reason, namely the necessity to take limits
with respect to certain infinitesimals (in particular $\zeta$ will be
treated as such), we shall use the 
approach presented in Sections~\ref{sect:limimagedim0} and
\ref{sect:limimage}.  

The main results of this section are as follows.
\begin{theorem}\label{th:pieces}
Let $\zeta$ be a regular value of $p(Q(X))$ and $p(Y)$, and
the level set $V=Z(p(Q(X))-\zeta,\K^n)$  be bounded over $\K$.
Further, let the matrices $\overl{H}_j$'s be in 
$r$-general position with respect to $V$. Then 
one can construct a covering of the set of critical
points $V_c\subseteq V$ of the projection map $X\mapsto X_1$
by semialgebraic sets $V_c(U,W)$ indexed by invertible submatrices
of $\sum_{j} p_j(Q(X))H_j$ with row sets
$U\subseteq\{2,\dots,n\}$ and column sets 
$W\subset\{1,\dots,n\}$ that
satisfy $r\leq |U|=|W|\leq n-1$. For each such $W$,  
the polynomial mapping
\begin{equation}\label{eq:defphimap}
\begin{split}
\phi_W : \K^n & \rightarrow \K^{k+n-|W|}\\
X & \mapsto \binom{Q(X)}
 {X_{\overline{W}}}
\end{split}
\end{equation}
on each 
$\phi_W(V_c(U,W))$
has explicitly given inverse %$\phi^{-1}_{UW}.$
\begin{equation*}%\label{eq:phiUWinverse}
\begin{split}
\phi^{-1}_{UW} : \K^{k+n-|W|}& \rightarrow \K^n\\
\binom{Y}{T}& \mapsto \binom{\Theta_{UW}(p_1(Y),\dots,p_k(Y),T)}{T},
\end{split}
\end{equation*}
%The map $\phi^{-1}_{UW}$ is a vector of rational functions
%in the $p_j$'s and $T=(T_1,\dots,T_{n-1-|W|}),$
where $\Theta_{UW}$ is a vector of rational functions
in $p_j(Y)$'s and $T,$  
all with the same denominator. The degrees of the latter and of the 
numerators are at most $|W|$.\\
The set $\phi(V_c(U,W))$ is defined by $p(Y)=\zeta$ and
$(n-|W|)^2+k+1+n-r$ polynomial 
equations and one un-equation in the $p_j$'s, $Y,$ and $T,$
 of degree $O(|W|).$ 
In total there are at most $n^{O(n-r)}$ pieces. 
\end{theorem}

\begin{remark}
The statement of Theorem~\ref{th:pieces} holds un-amended   
in the more general setting of $p$ being a differentiable
function, and ``semialgebraic set'' replaced by %``constructible variety''
``joint zeros of equations and non-zeros of inequations''.
The proof we give goes through in this case, too.
We chose to remain in semialgebraic setting for the sake
of clarity only.
\end{remark}

The following summarizes, in the setting of the paper,
 the necessary complexity estimates for 
the procedure outlined in the course of proving Theorem~\ref{th:pieces}.
\begin{proposition}\label{prop:pieces}
In notation of Theorem~\ref{th:pieces},
let $\K=\Le$, with $\eps=(\eps_1,\dots,\eps_\ell)$ infinitesimals,
$\ell\geq 1$,  
$D=D_\LL[\eps]$, and $p\in D[Y]$ of degree $d$ in $Y$.
Let the degrees of $Q$ and $p$ in $\eps$ be at most $d'$.
The map $\phi^{-1}_{UW}$  and the definition for
$\phi(V_c(U,W))$ can be computed within $(n(d+d'))^{O(\ell+k+n-r)}$ 
arithmetic operations in $D_\LL$.
In the case $D_\LL=\Z$, the bitsizes of them and the intermediate
data are bounded by the bitsize of $Q$ and $p$ times
a polynomial in $n,$ $k,$ $\log d,$ $\log (1+d'),$ and $\ell.$ 
\end{proposition}

In the course of the proof of
Theorem~\ref{th:pieces} we construct the explicit
maps $\phi^{-1}_{UW}$ and semialgebraic definitions for $V_c(U,W)$.
Equations \refeq{defVc} can be written as
$$\frac{\partial p(Q(X))}{\partial X_i}(X)= \sum_{j=1}^k
p_j(Q(X))e_i^T({H}_j X + b_j)=0,\qquad 2\leq i\leq k,$$
where $e_i$ is the standard $i$-th coordinate vector.
Thus, denoting
$$\Phi(Y)=\sum_{j=1}^k p_j(Y)\overl{H}_j,\qquad 
b(Y)=-\sum_{j=1}^k p_j(Y)\overl{b}_j,$$
where $\overl{b_j}$ denotes $b_j$ with the first coordinate
removed, one can write \refeq{defVc} in the matrix form, as follows.
\beq{Psi-linsys}
\Phi(Q(X))X=b(Q(X)).
\eeq
At this point we would like to give an outline of the remainder of the
proof of Theorem~\ref{th:pieces}.
We will compute a set of solutions of the system of 
equations \refeq{defV}-\refeq{Psi-linsys} that intersects each 
connected component of $V$. Doing this in the standard way would take
an exponential in $n$ number of operations, and this is exactly what
we want to avoid.

The structure of the equations \refeq{defV}-\refeq{Psi-linsys} 
suggests the substitution $Y=Q(X)$. It turns \refeq{Psi-linsys} 
into a system of linear equations $\Phi(Q(Y))X=b(Y)$ 
with respect to $X$. We cannot simply ``invert'' $\Phi(Q(Y))$, 
as it need not be of full rank. However, $\rank{\Phi(Q(Y))}$ will always
be at least $r$, allowing us to ``split'' the solving into
$n^{O(n-r)}$ cases, one for each maximal invertible submatrix
(parametrized by $U$ and $W$), that
will be inverted. This gives (for each such case) rational expressions
$\phi_{UW}^{-1}$ 
for $r$ of the $X$'s in terms of $Y$ and the remaining $n-r$ of $X$'s
(that we will denote $T$), 
as well as giving extra (in)equations involving $Y$ and $T$.
The latter define the sets $V_c(U,W)$, essentially completing the 
proof.

We proceed with the detailed proof now, preparing the ground for introducting
in \refeq{phiUWinverse} the variables $Y$.
The $r$-general position assumption implies
$$\rank{\Phi(Q(x))}\geq r\quad\text{for any}\quad x\in V.$$
Thus there are at most $n^{O(n-r)}$ maximal by inclusion 
invertible submatrices of $\Phi(Q(X)).$
Indeed, there are at most $s(r)=(\sum_{m=r}^{n}\binom{n}{m})^2$ 
of them, by counting 
number of pairs $(U,W)$ of subsets $U\subseteq\{2,\dots,n\},$ 
$W\subset\{1,\dots,n\}$ satisfying $r\leq |U|=|W|.$
If $r\leq n/2$ then   
$s(r)<2^{O(n)}<n^{O(n-r)}.$ Otherwise one has
$s(r)<(n-r)^2\binom{n}{n-r}^2<n^{O(n-r)},$ again as required.

As well, at least one $\Phi(Q(X))_{UW}$ will be invertible.  Hence
\begin{equation}\label{eq:defVcUW}
\begin{split}
\det \Phi(Q(X))_{U'W'}&=0\quad\text{for all}\quad |U'|=|W'|=|U|+1\\
& U\subset U'\subseteq \{2,\dots,n\},\ W\subset W'\subset
\{1,\dots,n\},\\
\det \Phi(Q(X))_{UW}&\neq 0.
\end{split}
\end{equation}
Noting that \refeq{defVcUW} implies that $\det \Phi(Q(X))_{U'W'}=0$ for
all $U'$ and $W'$ of size bigger than $|U|$, one obtains the following.
\begin{lemma}\label{lem:VcUW}
\beq{Vccover}
V_c=\bigcup_{\substack{U\subseteq\{2,\dots,n\}\\
W\subset\{1,\dots,n\}\\
r\leq |U|=|W|\leq n-1}} V_c(U,W),
\eeq
where $V_c(U,W)$ is defined by the equations \refeq{defV}-\refeq{defVc}
and by \refeq{defVcUW}. 
The number of elements in the union in 
\refeq{Vccover} is at most $n^{O(n-r)}.$ 
\end{lemma}
Without loss in generality
$U=\{2,\dots,r+1\}$, $W=\{1,\dots,r\}$.
Invert
the matrix  $\Psi=\Phi(Q(X))_{UW}$  using the Cramer rule:
\beq{Cramer}
\Psi^{-1}_{ij}=\frac{(-1)^{i+j}\det \Psi(i,j)}{\det \Psi},
\quad\text{for}\quad 1\leq i,j\leq r,
\eeq
where $\Psi(i,j)$ is the matrix obtained from $\Psi$ by removing
$i$-th row and $j$-th column.
Then the system $\Phi(Q(X))X=b(Q(X))$ 
can be rewritten in the block form as
\beq{blmat}
\begin{pmatrix}
\Psi&\Phi_{U,\overline{W}}\\ 
\Phi_{\overline{U},W}&\Phi_{\overline{UW}}
\end{pmatrix}
\begin{bmatrix} X_W\\ X_{\overline{W}} \end{bmatrix}=
\begin{bmatrix} b_U\\ b_W \end{bmatrix},
\eeq
where the common ``$(Q(X))$'' is dropped for the sake of readability.
Applying 
$\bigl( \begin{smallmatrix}
\Psi^{-1} & 0\\-\Phi_{\overline{U},W}\Psi^{-1} & I
\end{smallmatrix}\bigr)$
to both sides of \refeq{blmat}, one obtains
\beq{blmat-exp}
\begin{pmatrix}
I & \Psi^{-1}\Phi_{U,\overline{W}}\\ 0&0
\end{pmatrix}
\begin{bmatrix} X_W\\ X_{\overline{W}} \end{bmatrix}=
\begin{bmatrix} \Psi^{-1} b_U\\ 
b_W-\Phi_{\overline{U},W}\Psi^{-1}b_U \end{bmatrix}.
\eeq
Thus the following, together with 
\refeq{defVcUW} and \refeq{defV}, provides another definition of
$V_c(U,W)$.
\begin{align}\label{eq:defVcUW1}
X_W&=\Phi(Q(X))_{UW}^{-1}\cdot\left(b(Q(X))_U-
\Phi(Q(X))_{U\overline{W}} X_{\overline{W}} \right),\\ 
\label{eq:defVcUW2}
b(Q(X))_{\overline{U}}&=\Phi(Q(X))_{\overline{U}W}
\Phi(Q(X))_{UW}^{-1}\cdot b(Q(X))_U.
\end{align}
Observe that in the latter definition of $V_c(U,W)$
the only place where $X_W$ appears other than as an argument to $Q$
is the left-hand side of \refeq{defVcUW1}. 
Set up the mapping $\phi^{-1}_{UW}$ as follows.
\begin{equation}\label{eq:phiUWinverse}
\begin{split}
\phi^{-1}_{UW} : \K^{k+n-r}& \rightarrow \K^n\\
\binom{Y}{T}& \mapsto 
\binom{\Phi(Y)_{UW}^{-1}\cdot(b(Y)_U-
\Phi(Y)_{U\overline{W}} T )}{T}.
\end{split}
\end{equation}
%Comparing \refeq{defVcUW1} and \refeq{phiUWinverse}
One establishes
that $\phi^{-1}_{UW}$ acts as claimed in the statement of the theorem.
\begin{lemma}\label{lem:phiUWinverse}
The mapping $\phi_W$ restricted onto $V_c(U,W)$ has inverse
$\phi^{-1}_{UW}$, that is, 
$$\phi^{-1}_{UW}(\phi_W(V_c(U,W)))=V_c(U,W).$$
\end{lemma}
\begin{proof}
We have to check that 
$\phi^{-1}_{UW}(\phi_W(x))=x$ for any $x\in V_c(U,W)$.
As neither $\phi_W$ nor $\phi^{-1}_{UW}$ change $x_{\overline{W}}$, in view of
\refeq{phiUWinverse} it suffices to check that
$\Phi(Q(x))^{-1}_{UW}(b(Q(x))_U-\Phi(Q(x))_{U\overline{W}}x_{\overline{W}})
=x_W.$ But the latter holds as it is nothing but 
\refeq{defVcUW1}, a part of a definition for $V_c(U,W)$.
\end{proof}

We proceed to write down an explicit definition for  
$\phi_W(V_c(U,W))$ in terms of variables $Y$ and $T$ used in
\refeq{phiUWinverse}.
Denoting $\Omega=\det \Phi(Y)_{UW},$ one has the following
polynomial equations 
\begin{align}
p(Y)&=\zeta\label{eq:defVcUWim1} \\
\Omega^2\, Y&=\Omega^2\, Q(\phi_{UW}^{-1}(Y,T)) \label{eq:defVcUWim2}\\
\Omega\, b(Y)_{\overline{U}}&=\Omega\,
  \Phi(Y)_{\overline{U},W}\Phi(Y)_{UW}^{-1}\cdot b(Y)_U,
\label{eq:defVcUWim3}
\end{align} 
where puzzlingly looking multiplication of both sides of 
\refeq{defVcUWim2} and \refeq{defVcUWim3} by $\Omega$ clears
denominators coming from \refeq{Cramer} in the right-hand sides, and
\begin{equation}\label{eq:defVcUWim4}
\begin{split}
\det \Phi(Y)_{U'W'}&=0\quad\text{for all}\quad |U'|=|W'|=|U|+1\\
& U\subset U'\subseteq \{2,\dots,n\},\ W\subset W'\subseteq 
\{1,\dots,n\},\\
\det \Phi(Y)_{UW}&\neq 0.
\end{split}
\end{equation}
Apart from \refeq{defVcUWim2}, that ``bootstraps'' $Q$,
these (in)equations already appeared above, with $Y$ substituted
for $Q(X)$ and $T$ substituted for $X_{\overline{W}}$.
\begin{lemma}\label{lem:defpieceim}
The relations 
\refeq{defVcUWim1}-\refeq{defVcUWim4} provide a semialgebraic definition
of\\ $\phi_W(V_c(U,W)).$ 
\end{lemma}
\begin{proof}
Let $(y,t)$ belong to the semialgebraic set defined by
\refeq{defVcUWim1}-\refeq{defVcUWim4},
and $x=\phi^{-1}_{UW}(y,t)$.
We shall check that $x\in V_c(U,W)$.
Due to \refeq{defVcUWim2}, the equation \refeq{defV} holds for
$X=x$. Similarly, the remaining (in)equations 
\refeq{defVcUW}, \refeq{defVcUW1}-\refeq{defVcUW2} 
defining $V_c(U,W)$ hold for $X=x$.

By inspection, any $x\in V_c(U,W)$ gives rise to
$(y,t)=\phi_W(x)$ in the set defined by
\refeq{defVcUWim1}-\refeq{defVcUWim4}.
\end{proof}

The entries of the matrix $\Phi(Y)$ are 
linear polynomials in $p_1,\dots,p_k$.  Thus
the determinants of its $m\times m$-submatrices, that
come via \refeq{Cramer} into the definition of $V_c(U,W)$
by \refeq{defVcUWim1}-\refeq{defVcUWim4}, will be
polynomials of degree at most $m$. 
Similar straightforward degree counts complete 
the proof of Theorem~\ref{th:pieces}.

\medskip

We proceed to prove Proposition~\ref{prop:pieces}.
The only nontrivial part concerns the complexity of 
computing the map $\phi^{-1}(U,W)$ using the
Cramer rule  \refeq{Cramer}. 

To count the number of arithmetic
operations in $D_L$ required to compute
the determinants, 
one can either slightly extend \cite[(2.8)]{Bar93}, 
or refer to \cite[Alg.~8.38, Rem.~8.39(b)]{BPR03}, 
to obtain that computing the determinant of a submatrix
of $\Phi(Y)$ can be done in $n^{O(1)}$ arithmetic operations in
$D_L[\eps,Y]$, and then refer to \cite[Alg.~8.38]{BPR03}. 

Using the latter source, one sees that
for $D_L=\Z$, the bitsizes in the answer and in the
intermediate data will be bounded by $(\tau+\log n)n+(k+1)\log
(n(d+d')+1)$, with $\tau$ a bound on the bitsize of coefficients in the
entries of $\Phi(Y)$. Noting that $\tau$ is bounded by 
$\log d$ times the bitsize of $p$ and $Q$ completes the computation of
the bitsize bound for determinants.
The remainder of the proof is a straightforward
use of the complexity analysis of arithmetic operations in 
polynomial rings in \cite[Algs.~8.8,~8.10]{BPR03}.

\section{Proof of Theorem~\ref{thm:mainthm_pQ}}\label{sect:mainproof}
Let $\eps_0>0$ be an infinitesimal over $\K$. 
We use it to deform $Z=Z(p(Q(X)),\K^n)$ so that it becomes bounded.
(\emph{A priori} this deformation is not necessary if $Z$ is 
known to be bounded in the first place.)
We can assume that $p(Y)\geq 0$ for all $Y$, otherwise
we can replace $p$ by $p^2.$
Introduce extra variables $X_0$ and $Y_0$, and  set 
$$\tp(Y)=Y_0^2+p(Y).$$

Set $$Q_0(X)=1-\eps_0^2\sum_{i=0}^n X_i^2$$
and abuse slightly the notation by setting $Q=(Q_0,Q_1,\dots,Q_k).$
Further, set $\tk=k+1$ and $\tn=n+1$.
Then for $\hat{Z}=Z(\tp(Q(X)),\K\langle\eps_0\rangle^\tn)$ one sees,
by using \cite[Prop.~11.47]{BPR03}, that
the projection to the last $n$ coordinates of any set $S$ 
meeting every connected component of $\hat{Z}$ meets every
connected component of $Z(p(Q(X)),\K\langle\eps_0\rangle^n).$
Assuming one can compute $S$ as a set of univariate representations
in $D[\eps_0,T],$ one then can use \emph{restricted elimination} 
\cite[Alg.~12.43]{BPR03} to replace $\eps_0$ by a sufficiently small
element of the field of fractions of $D$ to obtain univariate
representations of points of $Z.$ 

Next, we deform $Q$ by defining $\tQ(t) : \K^\tn\rightarrow\K[t]^\tk$
as follows 
\beq{tQdef}
\tQ_{j}(t,X) = Q_j(X) + \frac{t}{2}X^T \diag{1^j,2^j,\dots,\tn^j}X,\qquad
0\leq j\leq k.
\eeq 
Obviously $\tQ(t)$ defines, as well, a quadratic map 
$\F^\tn\rightarrow\F^\tk$ for any field 
$\F\supseteq \K\langle t\rangle.$ 
The following lemma states that the Hessians of 
the $\tQ_{j}(t)$'s are in general position, 
in sense that their nonzero linear combinations
(that  
will be the matrices $A(Y)$ mentioned in the discussion
in the beginning of this section) are of maximal possible rank.
The statement 
is similar to \cite[(3.6)]{Bar93}.
\begin{lemma}\label{lem:rkbound}
Let the matrix $A(Y,T)$ with the entries in $\K[Y,T]$ be defined by 
$$A(Y,T)=\sum_{j=1}^{k} Y_{j}(H_j + T\, \diag{1^{j-1},2^{j-1},\dots,n^{j-1}}).$$
Let $t$ be transcendental over $\CK.$ Then 
for any field $\F\supseteq \K\langle t\rangle$  
and $0\neq y\in \overline\F^{k}$, 
the rank of $A(y,t)$ is at least $n-k+1$. \\
There exists $0<\iota'\in\K$ such that for any $\iota\in\K$ satisfying
$0<\iota<\iota'$ and $0\neq y\in\CK$ the rank of
$A(y,\iota)$ is at least $n-k+1$.
\end{lemma}
\begin{proof}
%Denote by $B_{UW}$ the submatrix of a matrix $B$ with
%rows in $U$ and columns in $W$, where $U,W\subseteq\{1,\dots,n\}$.
Consider $$B=B(Y,T,\mu)=\sum_{j=1}^{k} 
Y_{j}(\mu H_j + T\diag{1^{j-1},2^{j-1},\dots,n^{j-1}})$$
and the homogeneous, with respect to $Y$, as well as with respect to 
$\{\mu,T\}$, 
ideal $J=(\det{B_{UW}}\mid U,W\subset\{1,\dots,n\},\ |U|=|W|=n-k\})$
in the ring $D[Y,T,\mu]\subset\CK[Y,T,\mu]$.

Note that every $(y^*,t^*)\neq 0$ satisfying 
$\rank{A(y^*,t^*)}<n-k+1$ gives rise to 
$0\neq (y^*,t^*,1)\in Z(J)=Z(J,\CK^{k+2}).$ 
Vice versa, $0\neq (y^*,t^*,1)\in Z(J)$
obviously implies $\rank{A(y^*,z^*)}<n-k+1$.

The idea is now to show that there exists a nonzero polynomial
$f(T)\in\CK[T]$ such that $f(t)=0$
for all $(y^*,t)\neq 0$ with $\rank{A(y^*,t)}<n-k+1$.
As $t$ is transcendental over $\K$,
that is, it cannot be a root of a polynomial in
$\CK[T]$, this will imply the statement of the lemma.

The ideal $J'=(J:(Y)^\infty)\cap\CK[T,\mu]$, obtained by 
``projectively'' eliminating $Y$ from $J$,
is homogeneous.
By the ``Main Theorem of Elimination Theory'', see e.g.
\cite[Theorem~14.1]{Eis95}, the image of $Z(J)$ 
under the corresponding projection
is Zariski-closed. Hence it coincides with $Z(J')=Z(J',\CK^2)$.
Moreover, $J'$ is nonempty, as $Z(J)$, and hence $Z(J')$, do not contain
elements with $\mu=0$ and $T=1$, as follows immediately from the properties
of the diagonal matrices
$\diag{1^{j},2^{j},\dots,n^{j}}.$ Thus $J'$ contains a homogeneous polynomial 
$f(T,\mu)\in\CK[T,\mu]$
that is not divisible by $\mu$. 

Hence any $t$ for which there exists $y^*\neq 0$ satisfying
$\rank{A(y^*,t)}<n-k+1$, satisfies $f(t,1)=0$, that is, $t$ is
algebraic over $\CK,$ a contradiction showing the first part of the lemma.
To obtain the second part, observe that $f(T,1)$ vanishes only on finitely 
many elements of $\K$. Choose $\iota'$ to be the closest to $0$ root
of $f(T,1)$ among the positive elements of $\K$, if such a root exists.
Otherwise choose $\iota'=1$.
\end{proof}

Next, we shall perturb $\tp(\tQ(X))$ so that 0 is not its
(and neither that of $\tp(Y)$) critical value by subtracting
an appropriate constant $\tau$ from it. (Such $\tau$ is called a
\emph{regular value} of $\tp(\tQ(X))$ and of $\tp(Y)$). We will 
talk about the $\tau$-level set of $\tp(\tQ(X))$, that is just
$Z(\tp(\tQ(X))-\tau,\K^n)$. The following is an immediate consequence
of the semialgebraic Sard's theorem \cite[Thm.~9.6.2]{BCR98},
\cite[Thm.~5.57]{BPR03}.
\begin{lemma}\label{lem:pcrit}
Let $\F$ be a real closed field and $\tau$ a transcendental
infinitesimal 
over $\F$
(respectively, $\tau>0$ a sufficiently close to 0 element of $\F$).
Then $\tau$ (and any $\iota$ satisfying $0<\iota<\tau$) 
is a regular value of any nonzero $f(Y)\in\F[Y].$
In particular, provided that $\F$ contains the field
generated by the coefficients of $p$ and $Q$, one has that 
$\tau$ (and any $\iota$ as above) 
is a regular value of $\tp(\tQ(X))$ and of $\tp(Y).$ \qed
\end{lemma}

Let $\eps_0\gg\eps_1\gg\eps_2>0$ be two more extra 
infinitesimals over $\K$, 
and denote $\tQ=\tilde{Q}(\eps_2).$ We deform $\hat{Z}$ as follows:
$$\tilde{Z}=Z(\tp(\tQ(X))-\eps_1,
   \K\langle\eps_0,\eps_1,\eps_2\rangle^\tn).$$ 
At this point we are ready to use the tool from
Section~\ref{sect:pieces}, where it is described 
in slightly greater generality. According to 
Theorem~\ref{th:pieces} we have a covering of the set $V_c$ of the critical
points of $X\mapsto X_0$ on $\tilde{Z}$
by $n^{O(k)}$ semialgebraic sets $V_c(U,W)$.
Moreover, Theorem~\ref{th:pieces} gives us for each $V_c(U,W)$ 
an isomorphism $\phi_W$ (given by polynomials in $D[X]$ of degree at most $2d$) 
so that 
$\phi_W(V_c(U,W))\subseteq\K\langle\eps_0,\eps_1,\eps_2\rangle^{O(k)}$,
as well as its inverse $\phi^{-1}_{UW}$ (given by rational functions,
with common denominator, 
with coefficients in $D[\eps]$, of degrees at most $\tn$). 
By Proposition~\ref{prop:pieces},
this data can be computed by $(n(d+d'))^{O(k)}$ arithmetic
operations in $D$.

The sets $\phi_W(V_c(U,W))$ and $V_c(U,W)$ are both defined by equations
and one inequation $\Lambda\neq 0$, with
$\Lambda\in D[\eps][Y]$ (respectively, $\Lambda\in D[\eps][X]$). 
By adding one extra variable
as in the
beginning of Section~\ref{sect:limimage} we convert each of them into
a real algebraic set: 
add equation $Y_{\tk+1}\Lambda=1$ (respectively, $X_{\tn+1}\Lambda=1$)
and extend the maps 
$\phi_W$ and $\phi^{-1}_{UW}$ by ``ignoring'' these extra variables.
Apply to $Z_F:=\phi_W(V_c(U,W))$ % Z_0 -> Z_F
and $\Psi:=\phi^{-1}_{UW}$ the procedure of
Theorem~\ref{thm:imlim}.
It will produce a set
$R(U,W)$ of univariate representations of points intersecting each
connected component of $\lim_\eps V_c(U,W)$.

By the following lemma, the union of the $R(U,W)$'s over $U$, $W$ 
will intersect 
each connected component $C$ of $Z$, as 
by Proposition~\ref{prop:11.56} one has $C=\lim_{\eps_1,\eps_2} C_\eps$,
where $C_\eps$ is a connected component of $\tilde{Z}$, and 
$C_\eps$ intersects some $V_c(U,W)$.
\begin{lemma}
$\hat{Z}=\lim_{\eps_1,\eps_2} \tilde{Z}$.
\end{lemma}
\begin{proof}
Denote $\eps=(\eps_2,\eps_1)$.
As $\lim_\eps$ is a ring homomorphism from $\K\langle\eps\rangle_b$ to $\K,$
certainly $\lim_\eps \tilde{Z}\subseteq \hat{Z}.$ We shall show the reverse
inclusion. Let $x\in \hat{Z}.$ 
We find a point $\tilde{x}\in\tilde{Z}$ satisfying
$\lim_\eps\tilde{x}=x.$ Note that 
$\tp(\tQ(x))\in\eps_2\K$ and $\tp(\tilde{Q}(x))-\eps_1<0,$ 
as $\eps_1\gg \eps_2>0.$ On the other hand, as $\tp(Q(X))$ is not identically
0, for any $0<r\in\K$ the ball of radius $r$ around $x$ in $\K^\tn$
contains a point $y$ such that $\tp(Q(y))>0.$ 
As $x$ lies in the closure of the semialgebraic set $F_+$ defined by 
$\tp(Q(X))>0,$  there exists
a semialgebraic path $\gamma: [0,1]\rightarrow\K^\tn$ such that
$\gamma(0)=x$ and $\gamma((0,1])\subseteq F_+$, cf. Curve selection
lemma \cite[Thm.~3.19]{BPR03}. As the image  of a closed and
bounded semialgebraic set 
under a continuous semialgebraic function
on it is bounded, 
cf. \cite[Thm.~3.20]{BPR03}, $\gamma([0,1])$ is bounded over $\K.$

Let $\overline{\gamma}$ denote the extension of $\gamma$ to 
$\K\langle\eps\rangle$.
Then by \cite[Prop.~2.84]{BPR03} the set $\overline{\gamma}([0,1])$ is
bounded over $\K\langle\eps\rangle.$ 
By the semialgebraic intermediate value theorem 
\cite[Prop.~3.4]{BPR03} 
the set $\mathcal{I}(\tau_0)$ of all $\tau\in(0,\tau_0)\subset
\K\langle\eps\rangle$ satisfying
%\beq{intermed-tau}
%\[
$
\tp(\tQ(\overline{\gamma}(\tau)))=0
$
%\]
%\eeq
is a nonempty closed semialgebraic set.  
Choose $\tau$ in the closest to $0$ interval of 
$\mathcal{I}(\tau_0).$ Then $\lim_\eps\tau=0,$ 
as $\tau_0$ is arbitrary close to $0.$ As $\lim_\eps$ is a ring homomorphism,
we have $\tp(Q(\lim_\zeta\overline{\gamma}(\tau)))=0.$  

It remains to show that
$x=\lim_\eps\overline{\gamma}(\tau)$. Identify $\tau$ with the
corresponding (representative of the) germ of 
semialgebraic continuous functions on $\K_{>0}$
and think of $\overline\gamma(\tau)$ as of composition $\gamma\circ\tau$.
To complete the proof,
apply \cite[Lemma~3.21]{BPR03} that states that in this setting
($\gamma$ a semialgebraic continuous function on a closed bounded semialgebraic
set over $\K$ and $\tau$ an element of the extension of this set to 
$\K\langle\eps\rangle$)
one has $\gamma(\lim_\eps\tau)=\lim_\eps(\gamma\circ\tau).$ 
%It is easy to see by
%expanding $p(\tQ)$ in the powers of $\eps_2$ that 
%$p(\tQ(y))-\eps_1>0.$  
\end{proof}

At this point we obtained a set of univariate representations 
$u(\eps_0,T)\in D[\eps_0,T]^{n+3}$ (see \refeq{def:univar}) for
points in each connected component of $\hat{Z}$.
Now we get rid of the infinitesimal $\eps_0$.
Remove from $u$ the polynomial $g_1$ responsible for $X_0$-coordinate,
that is no longer needed, and
apply \cite[Alg.~12.46]{BPR03} (Removal of Infinitesimals), 
that consists of two steps.

The first step is running
the restricted elimination algorithm \cite[Alg.~12.43]{BPR03}
with the input consisting of the polynomial $f$ and the following polynomials:
\beq{reliminp}
f', f'', \dots, f^{(\deg(f)-1)},\quad 
g_0^{\deg(p(Q))} p(Q(\frac{g_{2}}{g_0},\dots,\frac{g_{n+1}}{g_0})).
\eeq
It outputs a finite set $\mathcal{S}\subset D[\eps_0]$ 
such that the degree of $f$, the number of roots of $f$ in $\K$,
the number of common roots of $f$ and $h$ in $\K$, and
the signs of $h$ at the roots of $f$, for all $h$
in \refeq{reliminp}, are fixed on each connected component of the
realization of any (realizable) sign condition on $\mathcal{S}$.

The second step computes, for each polynomial 
$$h(\eps_0)=h_\ell \eps_0^\ell+h_{\ell+1}\eps_0^{\ell+1}+
\dots+h_{\ell+\omega}\eps_0^{\ell+\omega}\in\mathcal{S},\qquad h_\ell\neq 0,$$ 
the Cauchy lower bound $\frac{|h_{\ell}|}{\sum_m |h_m|}$
on the absolute
value of its {\em nonzero} roots (see \cite[Lemma~10.3]{BPR03}) and 
substitutes the minimum
$\frac{a}{b}$, for $a,b\in D$, of these bounds for $\eps_0$.
The following remain unchanged upon substituting
$\eps_0=\frac{a}{b}$:
\begin{itemize}
\item the number of real roots of $f(\frac{a}{b},T)$ 
and their Thom encodings; 
\item the signs of the polynomials in \refeq{reliminp} at these
roots.  In particular the
Thom encodings of these roots will remain the same.
\end{itemize}
By construction, the set of points represented by the $u(\frac{a}{b},T)$'s
intersects each connected component of $Z$.

The number of arithmetic operations in $D$ for
these two steps is $(dn)^{O(k)}$, according to \cite[p.~462]{BPR03}.

\medskip

It remains to convert the $u(\frac{a}{b},T)=(f,g_0,g_2,\dots,g_{n+1})$'s 
into real univariate representations by
computing the Thom encodings $\sigma$ for each real root
of $f$ using \cite[Alg.~10.64]{BPR03} and \cite[Rem.~10.66]{BPR03}.
The complexity of this procedure is $(dn)^{O(k)}$, and
for the case $D=\Z$ the bitsizes of the intermediate data are bounded
by  $(dn)^{O(k)}$ times the bitsize of the input, by [loc.cit.].

This completes the proof of Theorem~\ref{thm:mainthm_pQ}.

\begin{acknowledge}
The second author is supported by NWO Grant 613.000.214.
Parts of the work were completed while
he held a position at THI/FB20 Informatik, University of  
Frankfurt (Germany) supported by DFG Grant SCHN-503/2-1.
As well, he was partially supported by 
Universit\'e de Rennes I (France) and 
the Mathematical Sciences Research Institute (USA).

The authors thank Saugata Basu, Marie-Fran{\c{c}}oise Roy, Nikolaj Vorobjov,
and two anonymous referees 
for useful remarks, and Michael Kettner for 
carefully reading a preliminary version of the text 
and tireless help in improving it. 
Sergey Fomin kindly provided references in
Remark~\ref{rem:b_i}.
\end{acknowledge}

\nocite{BPR96,GV88,GV96,Ren92}
\bibliography{qp}

\def\cprime{$'$} \def\cprime{$'$}
\begin{thebibliography}{10}

\bibitem{ABRW96}
M.-E. Alonso, E.~Becker, M.-F. Roy, and T.~W{\"o}rmann.
\newblock Zeros, multiplicities, and idempotents for zero-dimensional systems.
\newblock In {\em Algorithms in algebraic geometry and applications (Santander,
  1994)}, volume 143 of {\em Progr. Math.}, pages 1--15. Birkh\"auser, Basel,
  1996.

\bibitem{Bar93}
A.~I. Barvinok.
\newblock Feasibility testing for systems of real quadratic equations.
\newblock {\em Discrete Comput. Geom.}, 10(1):1--13, 1993.

\bibitem{Bar97}
A.~I. Barvinok.
\newblock On the {B}etti numbers of semialgebraic sets defined by few quadratic
  inequalities.
\newblock {\em Math. Z.}, 225(2):231--244, 1997.

\bibitem{BPR96}
S.~Basu, R.~Pollack, and M.-F. Roy.
\newblock On the combinatorial and algebraic complexity of quantifier
  elimination.
\newblock {\em J. ACM}, 43(6):1002--1045, 1996.

\bibitem{BPR03}
S.~Basu, R.~Pollack, and M.-F. Roy.
\newblock {\em Algorithms in Real Algebraic Geometry}.
\newblock Springer-Verlag, 2003.

\bibitem{BCR98}
J.~Bochnak, M.~Coste, and M.-F. Roy.
\newblock {\em Real algebraic geometry}.
\newblock Springer-Verlag, Berlin, 1998.
\newblock Translated from the 1987 French original, Revised by the authors.

\bibitem{Eis95}
D.~Eisenbud.
\newblock {\em Commutative Algebra {\small with a View Toward Algebraic
  Geometry}}.
\newblock Springer-Verlag, New York, 1995.

\bibitem{Gri88}
D.~Grigoriev.
\newblock Complexity of deciding {T}arski algebra.
\newblock {\em J. Symbolic Comput.}, 5(1-2):65--108, 1988.

\bibitem{GV96}
D.~Grigoriev and N.~Vorobjov.
\newblock Complexity lower bounds for computation trees with elementary
  transcendental function gates.
\newblock {\em Theoret. Comput. Sci.}, 157(2):185--214, 1996.

\bibitem{GV88}
D.~Grigoriev and N.~N. Vorobjov, Jr.
\newblock Solving systems of polynomial inequalities in subexponential time.
\newblock {\em J. Symbolic Comput.}, 5(1-2):37--64, 1988.

\bibitem{MR45:9480}
L.~E. Heindel.
\newblock Integer arithmetic algorithms for polynomial real zero determination.
\newblock {\em J. Assoc. Comput. Mach.}, 18:533--548, 1971.

\bibitem{LeVe}
U.~J.~J. Le~Verrier.
\newblock Sur les variations s\'eculaires des \'el\'ements elliptiques des sept
  plan\`etes principales: Mercure, v\'enus, la terre, mars, jupiter, saturne et
  uranus.
\newblock {\em J. Math. Pures Appl.}, 5:220--254, 1840.

\bibitem{Mac95}
I.~G. Macdonald.
\newblock {\em Symmetric functions and {H}all polynomials}.
\newblock Oxford Mathematical Monographs. The Clarendon Press, Oxford
  University Press, New York, second edition, 1995.
\newblock With contributions by A. Zelevinsky, Oxford Science Publications.

\bibitem{MR84g:20099}
E.~W. Mayr and A.~R. Meyer.
\newblock The complexity of the word problems for commutative semigroups and
  polynomial ideals.
\newblock {\em Adv. in Math.}, 46(3):305--329, 1982.

\bibitem{Ren92}
J.~Renegar.
\newblock On the computational complexity and geometry of the first-order
  theory of the reals. {I}-{I}{I}{I}.
\newblock {\em J. Symbolic Comput.}, 13(3):255--352, 1992.

\bibitem{RRS00}
F.~Rouillier, M.-F. Roy, and M.~{Safey El Din}.
\newblock Finding at least one point in each connected component of a real
  algebraic set defined by a single equation.
\newblock {\em J. Complexity}, 16:716--750, 2000.

\bibitem{GaGe99}
J.~von~zur Gathen and J.~Gerhard.
\newblock {\em Modern computer algebra}.
\newblock Cambridge University Press, New York, 1999.

\end{thebibliography}
%\bibliography{../qp}
\bibliographystyle{abbrv}

\end{document}